\documentstyle[psfig]{mn2e}
\newcommand{\etal}{{et al.~}}

\newcommand{\kmsmpc}{\>{\rm km}\,{\rm s}^{-1}\,{\rm Mpc}^{-1}}

\newcommand{\Mpc}{\>{\rm Mpc}}

\newcommand{\Msun}{\>{\rm M_{\odot}}}

\newcommand{\Lsun}{\>{\rm L_{\odot}}}

\newcommand{\beq}{\begin{equation}}
\newcommand{\eeq}{\end{equation}}
\newcommand{\kpch}{\>{h^{-1}{\rm kpc}}}
\newcommand{\mpch}{\>h^{-1}{\rm {Mpc}}}

\newcommand{\rmd}{{\rm d}}

\newcommand{\msunh}{\>h^{-1}\rm M_\odot}

\newcommand{\apj}{ApJ}
\newcommand{\apjs}{ApJS}
\newcommand{\aj}{AJ}
\newcommand{\mnras}{MNRAS}

\newdimen\hssize
\newdimen\hdsize 
\begin{document}

%%%%%%%%%%%%%%%%%%%%%%%%%%%%%%%%%%%%%%%%%%%%%%%%%%%%%%%%%%%%%%%%%%%%%%%%%%%%%%%

\title[Weak Lensing by Galaxies in Groups and Clusters]
      {Weak Lensing by Galaxies in Groups and Clusters:
        I.--Theoretical Expectations}
\author[X. Yang et al.]
       {\parbox[t]{\textwidth}{
        Xiaohu Yang$^{1,6}$ \thanks{E-mail: xhyang@shao.ac.cn},
        H.J. Mo$^{2}$, Frank C. van den Bosch$^{3}$, Y.P. Jing$^{1,6}$,\\
        Simone M. Weinmann$^{4}$, M. Meneghetti$^{5,7}$}\\
        \vspace*{3pt} \\
       $^1$Shanghai Astronomical Observatory; the Partner Group of MPA,
            Nandan Road 80, Shanghai 200030, China \\
       $^2$Department of Astronomy, University of Massachusetts,
            Amherst MA 01003-9305, USA\\
       $^3$Max-Planck Institute for Astronomy, K\"onigstuhl 17, 
           D-69117 Heidelberg, Germany\\
       $^4$Institute for  Theoretical  Physics, University  of
           Zurich,  CH-8057, Zurich,  Switzerland\\  
       $^5$Zentrum f\"ur Astronomie, ITA, Universit\"at Heidelberg,
           Albert-\"Uberle-Str.~2, D-69120 Heidelberg, Germany \\
       $^6$Joint Institute for Galaxy and Cosmology (JOINGC) of SHAO
           and USTC \\
       $^7$INAF-Osservatorio Astronomico di Bologna, Via Ranzani 1, 40127, 
           Bologna}

%%%%%%%%%%%%%%%%%%%%%%%%%%%%%%%%%%%%%%%%%%%%%%%%%%%%%%%%%%%%%%%%%%%%%%%%%%%%%%%

\date{}

%\pubyear{2005}

\maketitle

\label{firstpage}

%%%%%%%%%%%%%%%%%%%%%%%%%%%%%%%%%%%%%%%%%%%%%%%%%%%%%%%%%%%%%%%%%%%%%%%%%%%%%%%

\begin{abstract}
  Galaxy-galaxy lensing is rapidly  becoming one of the most promising
  means  to accurately  measure  the average  relation between  galaxy
  properties and halo mass.  In order to obtain a signal of sufficient
  signal-to-noise, one needs to  stack many lens galaxies according to
  their  property of  interest, such  as luminosity  or stellar  mass. 
  Since such a stack consists  of both central and satellite galaxies,
  which contribute very different lensing signals, the resulting shear
  measurements  can   be  difficult   to  interpret.   In   the  past,
  galaxy-galaxy  lensing studies have  either completely  ignored this
  problem,  have applied  rough isolation  criteria in  an  attempt to
  preferentially select `central' galaxies, or have tried to model the
  contribution of  satellites explicitely. However, if one  is able to
  {\it a priori} split the  galaxy population in central and satellite
  galaxies, one  can measure  their lensing signals  separately.  This
  not only allows  a much cleaner measurement of  the relation between
  halo mass  and their  galaxy populations, but  also allows  a direct
  measurement of  the sub-halo  masses around satellite  galaxies.  In
  this paper, we  use a realistic mock galaxy  redshift survey to show
  that galaxy groups, properly selected from large galaxy surveys, can
  be used  to accurately split  the galaxy population in  centrals and
  satellites.   Stacking  the resulting  centrals  according to  their
  group  mass, estimated  from the  total group  luminosity,  allows a
  remarkably accurate  recovery of the masses and  density profiles of
  their  host   haloes.   In  addition,   stacking  the  corresponding
  satellite galaxies  according to  their projected distance  from the
  group center  yields a lensing signal  that can be  used to accurate
  measure the masses of both  sub-haloes and host haloes.  We conclude
  that an  application of galaxy-galaxy lensing  measurements to group
  catalogues  extracted from  large galaxy  redshift surveys  offers a
  unique  opportunity to accurately  constrain the  galaxy-dark matter
  connection.
\end{abstract}

%%%%%%%%%%%%%%%%%%%%%%%%%%%%%%%%%%%%%%%%%%%%%%%%%%%%%%%%%%%%%%%%%%%%%%%%%%%%%%%


%%%%%%%%%%%%%%%%%%%%%%%%%%%%%%%%%%%%%%%%%%%%%%%%%%%%%%%%%%%%%%%%%%%%%%%%%%%%%%%

\section{Introduction}

Understanding the  connection between galaxies and  dark matter haloes
is a major challenge in  modern astrophysics.  From the perspective of
cosmology, the  galaxy-dark matter connection is required  in order to
translate  observations   of  galaxy   clustering  in  terms   of  the
distribution  of  (dark)  matter.   From  the  perspective  of  galaxy
formation, it reveals  how halo mass impacts on  the properties of the
galaxies  that form  within them,  and thus  how the  various physical
processes that play  a role in galaxy formation scale  with halo mass. 
Consequently,  a   great  amount  of   effort  has  been   devoted  to
establishing  the  galaxy-halo  connection, either  through  numerical
simulations  (e.g., Katz,  Weinberg  \& Hernquist  1996; Fardal  \etal
2001;  Kay \etal 2002;  Springel \etal  2005; Springel  2005), through
semi-analytical modeling  (e.g., White \& Frenk  1991; Kauffmann \etal
1993, 2004; Mo \& Fukugita 1996;  Mo, Mao \& White 1999; Somerville \&
Primack 1999; Cole \etal 2000;  Benson \etal 2002; van den Bosch 2002;
Croton \etal  2006), or via  statistical approaches (e.g.,  Jing \etal
1998; Seljak  2000; White 2001;  Berlind \& Weinberg 2002;  Yang \etal
2003b; van den Bosch \etal 2003; 2005c; Zheng \etal 2005; Cooray 2005,
2006).   However, neither  of these  methods provides  a  {\it direct}
diagnostic of the galaxy-halo connection.

Direct measures of the dark  matter haloes around galaxies come either
from dynamical tracers or  from gravitational lensing. Galaxy rotation
curves and strong lensing, although extremely powerful, typically only
probe the  inner part of the  dark matter halo, and  are therefore not
well  suited to  determine the  total halo  mass.  The  only dynamical
tracers that probe the gravitational potential sufficiently far out to
allow for  an accurate mass estimate are  satellite galaxies (Zaritsky
\& White 1994; McKay \etal 2002; Prada \etal 2003; Brainerd \& Specian
2003; van den Bosch \etal  2004; Conroy \etal 2005). However, arguably
the  best method  to  directly  probe the  dark  matter haloes  around
galaxies is galaxy-galaxy lensing,  whereby the gravitational field of
lensing  galaxies induces  small tangential  shear distortions  in the
images of distant background  galaxies (e.g. Natarajan \& Kneib 1997).  
Unfortunately, since the weak
lensing signal around individual galaxies is too small to be detected,
one  can only  infer  ensemble-averaged properties:  by combining  the
signal  from  a large  number  of  lensing  galaxies one  obtains  the
galaxy-mass cross correlation function, which can be used to infer the
mass distribution  around galaxies, but  only in a statistical  sense. 
Note, however,  that a similar problem  hampers the satellite-dynamics
method,  where the  number of  satellite galaxies  of  individual host
galaxies is  too small to  allow for a  reliable estimate of  the halo
mass, and ensemble averaging has to be used as well.

Since  the galaxy-galaxy lensing  signal is  so weak,  and one  has to
carefully  correct for  a  number of  observational  effects, such  as
anisotropies  in the point-spread  function and  shear induced  by the
camera optics, it  took twelve years since the  first attempt by Tyson
\etal  (1984)  until the  first  detections  of galaxy-galaxy  lensing
(Brainerd  \etal 1996;  Griffiths  \etal 1996;  dell'Antonio \&  Tyson
1996).   In recent  years, however,  the progress  has  been enormous,
largely  due to  the  advent of  deep  galaxy surveys  with large  sky
coverage (e.g.   Fischer \etal 2000;  Wilson \etal 2001;  Smith \etal
2001; Mckay \etal 2001; Hoekstra \etal 2003, 2004, 2005; Sheldon \etal
2004; Kleinheinrich  \etal 2006;  Parker \etal 2005;  Mandelbaum \etal
2006a,b).

The  fact that  galaxy-galaxy  lensing only  yields ensemble  averaged
properties  complicates  the   interpretation.   Since  galaxies  with
different luminosities and with different morphologies are expected to
reside  in  haloes of  different  masses,  the  ensemble averages  are
complicated sums  over large  ranges in halo  mass.  Unless one  has a
prior  knowledge of the  relation between  galaxy properties  and halo
mass,  an unbiased  interpretation  of the  measurements is  basically
impossible.   For example,  in the  past  most studies  have used  the
observed lensing signal only to constrain the normalization of an {\it
  assumed} relation between galaxy luminosity, $L$, and halo mass, $M$
(e.g.,  Fisher  \etal  2000;  Smith  \etal 2001;  Wilson  \etal  2001;
Hoekstra \etal 2003).  When  redshift information regarding the lenses
and/or sources is available, one can somewhat improve the constraints.
For example, Hudson \etal (1998) were able to simultaneously constrain
the normalization  and the slope of  the $L$-$M$ relation,  due to the
fact  that they  had photometric  redshifts available  for  both their
lenses and their sources.   Nevertheless, their results clearly depend
on the assumed functional form (a power-law) of the $L$-$M$ relation.

Therefore, one of the ultimate goals in galaxy-galaxy lensing is to be
able  to  measure the  lensing  signals  for  galaxies with  different
intrinsic properties.  This  not only requires a very  large sample of
galaxies (both lenses and sources), but also redshifts for the lensing
galaxies,  in order  to  be able  to  (i) convert  angular sizes  into
physical sizes, (ii) compute the luminosities of the lenses, and (iii)
reduce uncertainties  in the geometry of the  lens-source system.  The
Sloan Digital  Sky Survey  (SDSS) has proven  to be ideally  suited to
make progress along such a  direction, and several measurements of the
galaxy-galaxy  lensing signal  as  function of  galaxy luminosity  and
morphological  type have already  been made  (e.g., McKay  \etal 2001;
Sheldon \etal 2004; Mandelbaum \etal 2006a).

There is one additional problem that complicates the interpretation of
these  lensing  signals,  namely  the  distinction  between  `central'
galaxies,  which reside  at  the center  of  a dark  matter halo,  and
`satellite' galaxies, which  are located on an orbit  around a central
galaxy. In the current  paradigm, these satellite galaxies are thought
to be  associated with dark  matter sub-haloes, which are  haloes that
reside and  orbit within  a larger virialized  dark matter  halo.  The
idea that  satellite galaxies  are related to  the population  of dark
matter sub-haloes  is consistent with both  observations and numerical
simulations (e.g.   Kravtsov \etal 2004; Natarajan \& Springel 2004; 
Natarajan \etal 2006; Vale \&  Ostriker 2004, 2006;
Conroy  \etal 2006;  Kang \etal  2005).  The  lensing signal  around a
satellite galaxy  does not only  reflect the mass distribution  of the
dark matter  sub-halo in  which it  is located, but  also that  of the
larger halo that  hosts the sub-halo (hereafter host  halo).  Thus one
expects  the  lensing  signal  from  a satellite  galaxy  to  be  very
different  from that  of a  central galaxy  (e.g., Appendix~B  in Yang
\etal 2003a).  In addition, the lensing signal from a satellite galaxy
not only depends on the masses of the sub-halo and host halo, but also
on the geometrical  orientation of these two entities  with respect to
each other.

Since a galaxy of a  given luminosity and/or morphological type can be
either a central galaxy or a satellite galaxy, even the lensing signal
from a sample of lensing  galaxies with a narrow range in luminosities
and morphological types  can be difficult to interpret.   In the past,
galaxy-galaxy  lensing  studies have  either  completely ignored  this
problem  (e.g., McKay  \etal 2001;  Hoekstra \etal  2003,  2004), have
applied  rough  isolation criteria  in  an  attempt to  preferentially
select `central' galaxies (e.g., Hoekstra \etal 2005; Mandelbaum \etal
2006b), or  have modeled  the contribution of  satellites explicitely,
using  either semi-analytical  models for  galaxy formation  (Guzik \&
Seljak  2001; Yang \etal  2003a) or  a model  for the  halo occupation
statistics (Guzik \& Seljak 2002; Mandelbaum \etal 2005b, 2006a). None 
of these approaches have taken into account the weak lensing signals 
around satellite galaxies at different halo-centric distances.  

In this paper we demonstrate that it is actually possible to measure the
lensing signal around satellite galaxies at different projected halo-centric
distances, provided that a well-defined sample of galaxy groups and clusters
is available to represent galaxy distribution in dark matter haloes.  This not
only allows a much cleaner measurement of the relation between halo mass and
their galaxy populations, but also allows a direct measurement of the sub-halo
masses around satellite galaxies.  Using realistic mock galaxy catalogues we
show that the halo-based group finder, recently developed by Yang \etal
(2005a), allows an accurate identification of central and satellite galaxies,
and that their corresponding lensing signals allow an accurate recovery of
their mean host and sub-halo masses, respectively.  Note also that in order to
perform such a measurement, a large spectroscopic sample is needed to group
lensing galaxies according to their common haloes.  Currently, the largest
sample of this kind is from the Sloan Digital Sky Survey (SDSS).  With about
1/3 of the eventual SDSS, significant galaxy-galaxy lensing signals have been
detected separately for three (seven) subsamples in galaxy luminosity and
colour (e.g. Sheldon et al. 2004; Mandelbaum et al. 2006a).  This suggests
that the present SDSS data may already be able to probe the lensing signals
around galaxies in broad bins of halo mass and halo-centric distances.  With
the completion of the SDSS, and with future deeper imaging surveys in the same
part of the sky as the SDSS, we anticipate that the effects we are considering
here can be studied.  In a forthcoming paper, we will construct shear maps from
realistic mock catalogues taking into account observational effects to test
the feasibility of the method we are proposing here.

This paper  is organized as follows.   In Section~\ref{sec_lensing} we
give a  detailed description  of the weak  lensing signal  produced by
host haloes and sub-haloes, separately.  in Section~\ref{sec_Nbody} we
use realistic  mock catalogues constructed from  N-body simulations to
demonstrate  how  to  split  the  galaxy  population  in  central  and
satellite galaxies, and how  their lensing signals allow a measurement
of  their corresponding  host and  sub-halo masses,  respectively. Our
conclusions are summarized in Section~\ref{sec_discussion}.

\section{Lensing by haloes and sub-haloes}
\label{sec_lensing}

Galaxy-galaxy lensing  measures the profiles of  the tangential shear,
$\gamma_t(R)$, azimuthally  averaged over a thin  annulus of projected
radius $R$ around a set  of lens galaxies. This observable quantity is
related to the mean projected surface mass density within the aperture
radius $R$ according to
\begin{equation}
\label{eq:GMCCF}
\gamma_t(R) \Sigma_{\rm crit} =  \Sigma(\leq R) - \Sigma(R) \equiv
\Delta\Sigma(R)\,.
\end{equation}
Here  $\Sigma(\leq  R)$  is  the  mean  surface  density  within  $R$,
$\Sigma(R)$ is the azimuthally averaged surface density at $R$, and
\begin{equation}
\Sigma_{\rm crit}= \frac{c^2}{4 \pi G}
      \frac{D_{\rm s}}{D_{\rm l} D_{\rm ls}}
\end{equation}
is the  critical density, which is  determined by the  geometry of the
lens-source system  (Miralda-Escud\'e 1991;  see Schneider 2005  for a
detailed review).  In the above equation, $D_{\rm l}$, $D_{\rm s}$ and
$D_{\rm ls}$  are the angular diameter  distances to the  lens, to the
source and  between the  lens and the  source, respectively.   Since a
uniform mass sheet, such as the mean density of the universe, does not
contribute to  $\Delta\Sigma$, it basically measures  the {\it excess}
surface density (hereafter ESD).

The mean  excess surface density around  a galaxy is  specified by the
line-of-sight  projection  of   the  galaxy-matter  cross  correlation
function, $\xi_{\rm gm}(r)$, so that
\begin{equation}
\label{sigatr}
\Sigma(R) = 2 \overline{\rho} \int_{R}^{\infty} \xi_{\rm gm}(r) 
{r \, \rmd r \over \sqrt{r^2 - R^2}}\,,
\end{equation}
and
\begin{equation}
\label{siginr}
\Sigma(\leq R) = \frac{4\overline{\rho}}{R^2} \int_0^R y\,\,dy\,
 \int_{y}^{\infty} \xi_{\rm gm}(r) {r \, \rmd r \over \sqrt{r^2 - y^2}}\,
\end{equation}
with $\overline{\rho}$ the average background density of the Universe.  Note
in both equations, we have omitted the contribution from the mean density of
the universe, as it does not contribute to the ESD.  As we will see below, it
is important to distinguish between the lensing signal due to host haloes
(those haloes that are not embedded in a larger virialized structure) and
sub-haloes (haloes embedded in a host halo).  In what follows we refer to
galaxies at the centers of host- and sub-haloes as central galaxies and
satellite galaxies, respectively.

In the halo  model, the dark matter distribution  consists entirely of
dark  matter haloes,  and the  galaxy-mass cross  correlation function
consists of four  terms: three one-halo terms, and  one two-halo term. 
The  first one-halo  term is  due to  the host  haloes  around central
galaxies.   Satellite  galaxies  contribute  two one-halo  terms:  one
describing the  density distribution of  the dark matter  sub-halo and
one describing the density distribution  of the host halo in which the
sub-halo is embedded.   This latter term also depends  on the relative
location  of the sub-halo  with respect  to the  center of  the parent
halo.  Finally,  the two-halo  term describes the  correlation between
the  lens galaxy  and  the  large scale  distribution  of dark  matter
haloes.  In this paper we focus only on the expected lensing signal at
small $R$,  and we  will therefore ignore  this two-halo term  in what
follows.

From the above  it is clear that central  and satellite galaxies yield
different lensing  signals (see Hudson  \etal 1998; Yang  \etal 2003a;
Guzik \&  Seljak 2002).  In the  past, various studies  have used halo
occupation statistics to model the total galaxy-mass cross correlation
function,  due to central  and satellite  galaxies combined  (Guzik \&
Seljak 2002; Mandelbaum \etal 2005b,  2006a; Yoo \etal 2005).  Here we
adopt a different approach and we investigate the $\Delta\Sigma(R)$ of
central and satellite  galaxies separately. In Section~\ref{sec_Nbody}
we  show that  with  a decent  galaxy  group finder  one can  identify
central  and  satellite  galaxies,  and probe  their  lensing  signals
separately.

\subsection{Density Distribution of Haloes and Sub-Haloes}
\label{sec:densprof}

As outlined above, the tangential shear due to galaxy-galaxy lensing can be
used to infer the galaxy-mass cross correlation function $\xi_{\rm gm}(r)$.
On small scales, $\xi_{\rm gm}(r) = \rho(r)/\overline{\rho} - 1$, with
$\rho(r)$ the azimuthally averaged density distribution around the lensing
galaxies.  This in turn reflects the density distribution of the host- and of
the sub-haloes surrounding central and satellite galaxies, respectively.  For
simplicity, we will not treat the baryonic masses of the lens galaxies as
separate mass components, but instead consider them included in the halo
components. As shown in Lin et al. (2006), the baryonic component can change
the concentration of the total mass profile on small scales. We do not expect
such effect to have an important impact on our results, because we are
focusing on lensing signals on scales $R\ga 50 \kpch$.

For the host haloes,  we adopt  the following  density  profile:
\begin{equation}
\label{rhonfw}
\rho(r)={\rho_0\over (r/r_c)^\alpha (1+r/r_c)^{3-\alpha}}\,,
\label{eq:rho}
\end{equation}
where
\begin{equation}
\rho_0={{\overline\rho}\Delta_{\rm vir}\over 3 I(c,\alpha)}\,;~~~~~
I(c,\alpha)\equiv {1\over c^3}\int_0^c {dx\over x^{\alpha-2}
(1+x)^{3-\alpha}} \;
\end{equation}
(see  e.g., Zhao  1996;  Jing  \& Suto  2000).   Note that  $\alpha=1$
corresponds  to  the  NFW  profile  (Navarro, Frenk  \&  White  1997);
$\alpha=1.5$  corresponds  to  the  profile proposed  by  Moore  \etal
(1999), and  $\alpha=0$ corresponds to  a density distribution  with a
constant density core.   Although we will mainly focus  on haloes with
$\alpha=1$,  we also  briefly discuss  the  impact of  changing  the
central  cusp  slope.   This  is   motivated  by  the  fact  that  (i)
observationally the exact  value of the inner slope  $\alpha$ is still
uncertain (e.g., Swaters \etal 2003; Simon \etal 2005; Sand \etal 2004;
Bartelmann \& Meneghetti 2004; Gentile \etal 2004, 2005; 
Meneghetti et al. 2005), (ii) numerical
simulations suggest that haloes may reveal a fair amount of scatter in
their central cusp slopes (e.g.,  Jing \& Suto 2000; Power \etal 2003;
Dahle  \etal  2003;  Navarro   \etal  2004),  and  (iii)  the  density
distribution  is interpreted  as including  the contribution  from the
baryons, so that it doesn't have  to be in perfect accord with that of
dark matter haloes.

For  a given value  of $\alpha$,  the density  distribution of  a dark
matter  host halo  is specified  by two  parameters,  a characteristic
density $\rho_0$ and a characteristic radius $r_c$. Alternatively, one
can parameterize the  halo by its mass $M  = (4\pi/3) \Delta_{\rm vir}
\,  {\overline\rho}  \,  r_{\rm  vir}^3$ and  concentration  parameter
$c=r_{\rm vir}/r_c$. Here $r_{\rm  vir}$ is the virial radius, defined
so  that   the  average  density   within  it  is   $\Delta_{\rm  vir}
{\overline\rho}$.   Throughout  this   paper  we   adopt  $\Delta_{\rm
  vir}=180$, and  we use  $M_h$ and  $M_s$ to refer  to the  masses of
host- and sub-haloes, respectively.

Numerical simulations have shown that halo concentration is correlated
with  halo mass.  Throughout  we adopt  the model  of Eke,  Navarro \&
Steinmetz (2001)  to model  the relation between  $c$ and $M$,  and we
assume  that it  is free  of scatter.   Therefore, the  entire density
distribution of a  dark matter (host) halo is  completely specified by
its mass alone.

For sub-haloes,  we follow Hayashi \etal (2003),  who, using numerical
simulations, found  that the density profiles of  stripped dark matter
sub-haloes can be written as
\begin{equation}
\label{eq:rhos}
\rho_s(r) = {f_t \over 1 + ({r/r_{t,{\rm eff}}})^3}\, \rho(r)\,.
\end{equation}
Here $f_t$  is a  dimensionless measure for  the reduction  in central
density,  and $r_{t,{\rm eff}}$  is an  `effective' tidal  radius that
describes  the  outer  cutoff  imposed  by  tides.   For  $f_t=1$  and
$r_{t,{\rm eff}} \gg r_c$, eq.~(\ref{eq:rhos}) reduces to the original
mass  profile $\rho(r)$  given in  eq.~(\ref{eq:rho}), i.e.,  the mass
profile  before the  sub-halo  was  accreted by  the  host halo.   The
parameters  $f_t$ and  $r_{t,{\rm eff}}$  are determined  by  the mass
fraction of the sub-halo that  remains bound, $f_m$.  Fitting the mass
profiles of  numerous stripped  dark matter sub-haloes,  Hayashi \etal
(2003) obtained the following  relations between $r_{t,{\rm eff}}$ (in
units of the characteristic radius of the sub-halo) and $f_m$:
\begin{equation}
\label{eq:rte}
\log [r_{t,{\rm eff}}/r_c] = 1.02 + 1.38 \, \log f_m + 
0.37 \, (\log f_m)^2 \,,
\end{equation}
and between $f_t$ and $f_m$:
\begin{equation}
\label{eq:ft}
\log f_t  = -0.007 +0.35 \log f_m + 0.39 (\log f_m)^2 + 0.23 (\log f_m)^3.
\end{equation}

Using  a large  cosmological  numerical simulation,  Gao \etal  (2004)
studied the radial dependence of the retained mass fraction $f_m$ of a
large sample of sub-haloes.  Using the results shown in their Fig.~15,
we obtain the following mean relation:
\begin{equation}
f_m=0.65 (r_{s}/r_{{\rm  vir},h})^{2/3}\,,
\label{eq:fm}
\end{equation}
where $r_{s}$ is  the distance of the sub-halo from  the center of the
host  halo, and $r_{{\rm  vir},h}$ is  the virial  radius of  the host
halo.  The combination of  Eqs. (\ref{eq:rhos}) - (\ref{eq:fm}) give a
model for the density profile of a sub-halo with a given original mass
located at a given distance from its host halo.

Note that the above model for the density distribution of the host and
sub-haloes is only approximate.  For example, host haloes of fixed mass show a
fair amount of scatter in halo concentrations, correlated with the halo
formation time (e.g., Wechsler et al.  2002; Zhao \etal 2003a;b; Lu \etal
2006), which we completely ignore. In addition, the retained mass fraction of
sub-haloes is assumed to depend only on the instantaneous location of the
sub-halo.  In reality, however, sub-haloes with the same $r_{s}$ can have very
different $f_m$, depending on their orbital eccentricities and their time
since being accreted by the host halo.  Furthermore, eq.~(\ref{eq:fm}) was
obtained for host haloes with masses $\sim 10^{14} h^{-1} M_{\odot}$, while we
assume that it holds for haloes of all masses.  Some of these shortcomings may
affect our results. For example, as shown in Mandelbaum et al. (2005b), the
scatter in the mass-luminosity relation can cause the derived mass to deviate
from the mean sample halo mass, and so the uncertainty in the mass model of
satellite galaxies may add uncertainty in our results.  Unfortunately, a much
more realistic model for the mass and density is not available at the present
time, and we have to live with such uncertainties here.

\begin{figure*}
\centerline{\psfig{figure=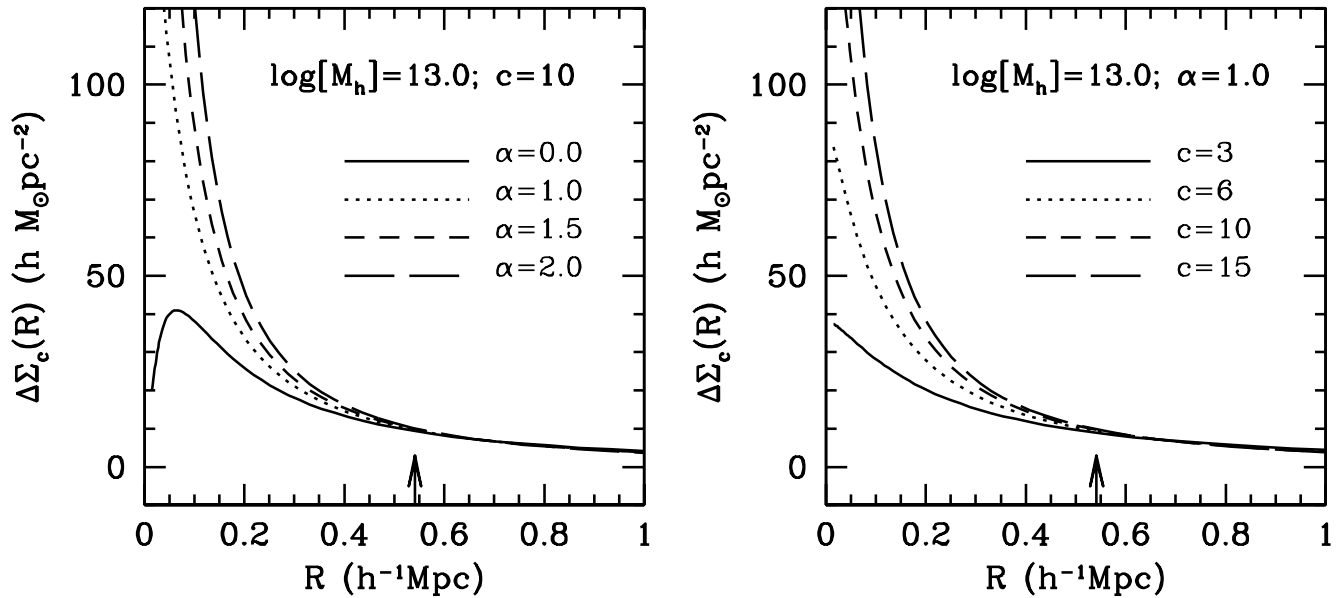,width=\hdsize}}
\caption{ The excess surface density (ESD) around central galaxies as  
  a function  of radius  $R$.  Results are  shown for  $10^{13} h^{-1}
  M_{\odot}$ dark matter haloes  with different density profiles.  The
  left-hand  panel   shows  model   predictions  for  haloes   with  a
  concentration  parameter $c=10$  but with  different values  for the
  central cusp slope, $\alpha$. The right-hand panel shows the ESDs of
  NFW  profiles  (i.e.,  $\alpha=1$)  with  different  concentration
  parameters, $c$.  Note that in the presence of realistic measurement
  errors,  there is a  significant degeneracy  between cusp  slope and
  halo concentration. The arrows in both panels indicate the virial 
  radius $r_{\rm vir}$ of a $10^{13} h^{-1}M_{\odot}$ dark matter halo.}
\label{fig:alpha}
\end{figure*}

\subsection{Central Galaxies}
\label{sec:central}

We  first consider  the lensing  signal around  central  galaxies.  At
sufficiently small $R$  we may simply replace $\overline{\rho}\xi_{\rm
  gm}(r)$  in eqs~(\ref{sigatr})  and~(\ref{siginr}) with  the density
distribution of  the host  halo, $\rho(r)$.  In  what follows,  we use
$\Delta\Sigma_c(R)$ to refer to the ESD of central galaxies.

For NFW profiles (i.e., $\alpha=1$) we can write (see Wright \& Brainerd
2000),
\begin{equation}\label{eq:SMDC_cen}
\Delta\Sigma_c(R) = {M \over 2 \pi r_c^2} I^{-1}(c,1) 
\left[g(R/r_c) - f(R/r_c)\right]\,,
\end{equation}
where
\begin{equation}
f(x) = \left\{ \begin{array}{ll}
\frac{1}{x^{2}-1}\left(1-\frac{{\ln \left({\frac{1+\sqrt{1-x^2}}{x}}\right)}}
{\sqrt{1-x^{2}}}\right) & \mbox{if $x<1$} \\
\frac{1}{3} & \mbox{if $x=1$}\\
\frac{1}{x^{2}-1}\left(1-\frac{{\rm atan} \left( \sqrt{x^2-1} \right) }
{\sqrt{x^{2}-1}}\right) & \mbox{if $x>1$}
\end{array}\right.
\end{equation}
and
\begin{equation}
g(x) = \left\{ \begin{array}{ll}
\frac{2}{x^{2}}\left(\ln (\frac{x}{2}) +
 \frac{{\ln \left( {\frac{1+\sqrt{1-x^2}}{x}}\right) }}{\sqrt{1-x^{2}}}\right) 
& \mbox{if $x<1$} \\
2+2\ln (\frac{1}{2}) & \mbox{if $x=1$} \\
\frac{2}{x^{2}}\left(\ln (\frac{x}{2}) + \frac{{\rm
atan}\left( \sqrt{x^2-1} \right) }{\sqrt{x^{2}-1}}\right) & \mbox{if $x>1$}
\end{array}\right.
\end{equation}

As an  illustration, Fig.~\ref{fig:alpha} shows  $\Delta\Sigma_c$ as a
function of  the projected  radius $R$ for  dark matter haloes  with a
mass $M_h = 10^{13}\msunh$. The left-hand panel depicts cases in which
the halo concentration is fixed at $c=10$, but in which the inner cusp
slope $\alpha$ changes from 0 to 2. The right-hand panel, on the other
hand, shows  results for NFW profiles ($\alpha=1$)  but with different
concentrations.  As  expected, for a  given $c$, a model  with smaller
$\alpha$ has  a shallower ESD  at small radii.   For the model  with a
constant  density core  ($\alpha=0$),  the  ESD goes  to  zero at  the
center, because  there is no  central density gradient that  can cause
image  distortions.  For  a given  $\alpha$, more  concentrated haloes
have a  more concentrated ESD, as  expected (see also  Guzik \& Seljak
2002).  Comparing the results shown  in the left and right panels, one
can  see  that there  is  a  degeneracy  between the  central  density
gradient and  the concentration  of a dark  matter halo, at  least for
realistic  observational errors  in  the tangential  shear. A  similar
degeneracy   also  hampers   a  unique   derivation  of   the  density
distribution of  dark matter haloes  from the rotation curves  of disk
galaxies (e.g., van den Bosch \etal 2000).
\begin{figure*}
\centerline{\psfig{figure=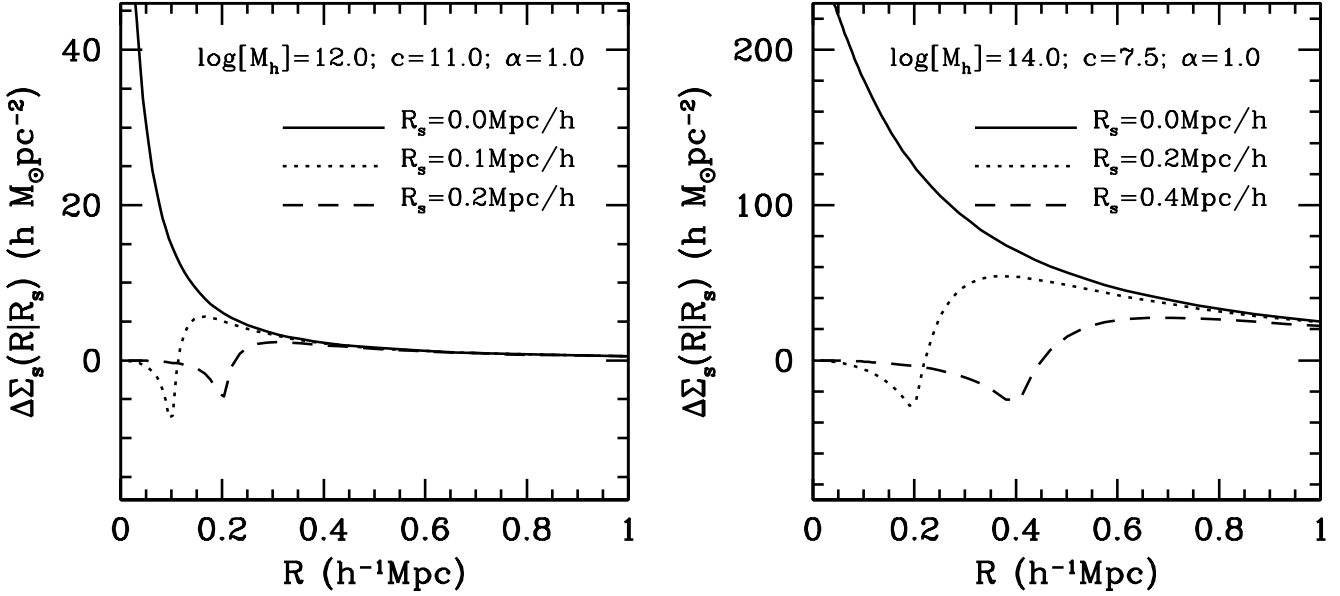,width=\hdsize}}
\caption{The excess surface density (ESD) around central galaxies
  (solid lines)  and satellite galaxies as  a function of  radius $R$. 
  Here it is  assumed that satellite galaxies are  not associated with
  dark matter sub-haloes; i.e., the ESD only reflects the contribution
  from the  host halo.  Different line styles  correspond to different
  halo-centric  distances of  the  satellite galaxies,  as indicated.  
  Panels of  the left and right  show the results for  NFW host haloes
  with       masses        of       $10^{12}h^{-1}M_{\odot}$       and
  $10^{14}h^{-1}M_{\odot}$, respectively.}
\label{fig:s0}
\end{figure*}

\subsection{Satellite Galaxies}
\label{sec:satellite}

The excess surface density around a satellite galaxy can be written as
\begin{equation}
\label{delsigsat}
\Delta\Sigma_s(R|R_s) = \Delta\Sigma_{s,{\rm sub}}(R) +
\Delta\Sigma_{s,{\rm host}}(R|R_s) 
\end{equation}
Here   $\Delta\Sigma_{s,{\rm   sub}}(R)$   and   $\Delta\Sigma_{s,{\rm
    host}}(R|R_s)$ are  the ESDs due  to the dark matter  sub-halo and
host halo,  respectively, and $R_s$ is the  projected distance between
the satellite  galaxy and the center  of its host  halo (throughout we
assume that a satellite resides at the center of its sub-halo).

The azimuthally  averaged, projected surface mass density  of the host
halo around a  satellite galaxy located at a  projected distance $R_s$
from the halo center is
\begin{equation}
\label{sigsat}
\Sigma_{s,{\rm host}}(R|R_s) = \frac {1}{2\pi} \int_{0}^{2\pi}
\Sigma\left(\sqrt{R_s^2 + R^2 + 2 R_s R \cos{\theta}}\right) 
\,{\rm d}\theta
\end{equation}
where $\Sigma (R)$ is the projected  density profile of the host halo. 
By   integrating~(\ref{sigsat})   from   $0$   to  $R$   one   obtains
$\Sigma_{s,{\rm    host}}(\le    R|R_s)$,    and    thus    the    ESD
$\Delta\Sigma_{s,{\rm host}}(R|R_s)$. The azimuthal averaging reflects
the fact that in order to obtain sufficient signal-to-noise to measure
the tangential shear one has to stack many satellite galaxies with the
same $R_s$. As long as  these have random orientation angles $\theta$,
the azimuthal averaging of eq.~(\ref{sigsat}) is appropriate.

To illustrate the ESD around satellite galaxies, we first show the
contribution from the host halo, by assuming that the satellite galaxy has no
corresponding sub-halo (i.e., $\Sigma_{s,{\rm sub}}(R)=0$).  Fig.~\ref{fig:s0}
shows the model predictions of $\Delta\Sigma_{s,{\rm host}}(R|R_s)$ around
satellite galaxies in NFW host haloes with masses of $10^{12}\msunh$ (left
panel) and $10^{14}\msunh$ (right panel), respectively.  Different line styles
refer to different halo-centric distances, as indicated. For comparison, we
also show the results for $R_s=0$ (solid lines), which are equivalent to the
ESDs around central galaxies.  Clearly, the ESD around satellite galaxies
(without sub-haloes) are very different from those around central galaxies.
They start with a value close to zero at $R=0$, decrease to a {\it negative}
minimum near $R_s$, and then increase rapidly with radius, eventually
approaching the ESD of central galaxies at $R\ga 3R_s$.  Note that here we are
measuring the ESD $\Delta\Sigma_{s,{\rm host}}(R|R_s)=\Sigma_{s,{\rm
    host}}(\le R|R_s)-\Sigma_{s,{\rm host}}(R|R_s)$ around satellite galaxies.
When $\Sigma_{s,{\rm host}}(R|R_s)$ reaches its maximum value at around $R_s$,
the ESD decreases to its negative minimum.  As expected, the overall amplitude
of $\Delta\Sigma_{s,{\rm host}}(R|R_s)$ at large radii is higher for more
massive haloes.  Thus, if the value of $R_s$ is known, the tangential shear
$\gamma_t(R)$ measured around satellite galaxies can be used to constrain the
mass distribution of their host haloes. We will come back to this in Section
~\ref{sec_GMCCF}.
\begin{figure*}
\centerline{\psfig{figure=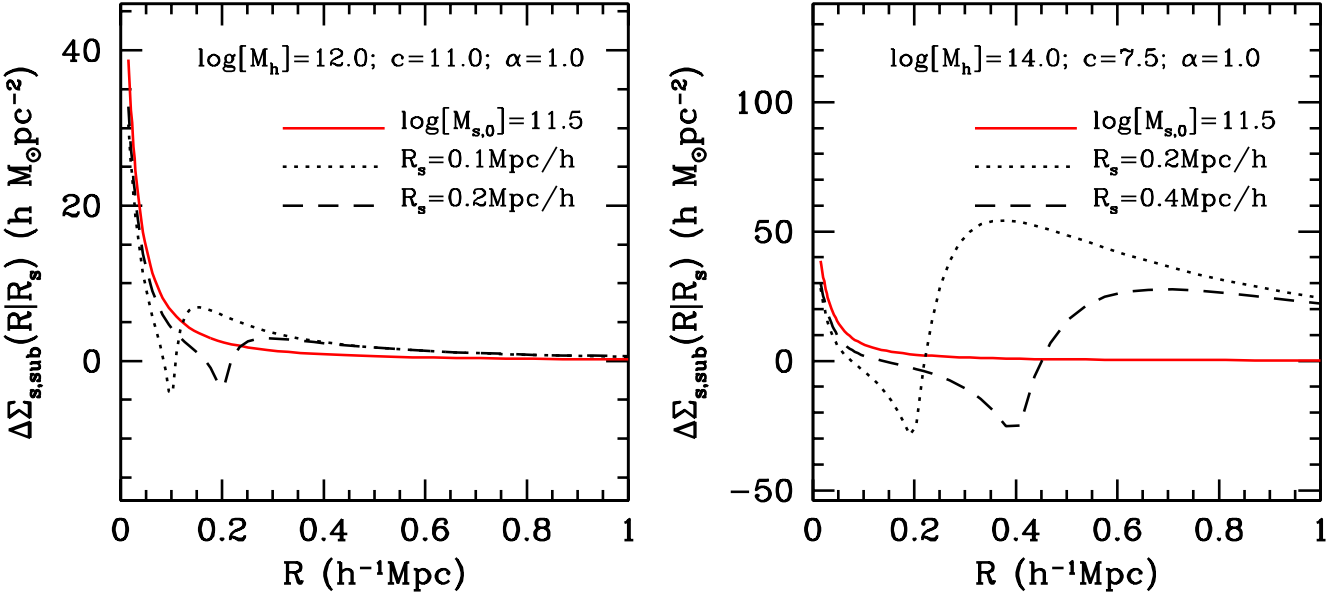,width=\hdsize}}
\caption{Same as Fig~\ref{fig:s0}, except that this time
  satellite galaxies are assumed to reside in sub-haloes whose mass is
  equal to $M_{\rm s,0}=10^{11.5}\msunh$ at the time of accretion. The
  solid  lines correspond  to the  ESD  of a  halo of  this mass  with
  $R_s=0$, and is shown for comparison.}
\label{fig:s11.5}
\end{figure*}

We now include the contribution of  the sub-halo. For this, we use the
sub-halo model  described in Section~\ref{sec:densprof},  which allows
us  to compute  $\Delta\Sigma_{s,{\rm sub}}(R)$  for a  given sub-halo
mass.  This  mass can be related to  the sub-halo mass at  the time of
accretion, using  eq.~(\ref{eq:fm}) and  the true distance  $r_{s}$ of
the sub-halo from the center of the host halo. In order to compute the
contribution   due    to   the   host    halo,   $\Delta\Sigma_{s,{\rm
    host}}(R|R_s)$,  one  also  needs  to  know $R_s$,  which  is  the
projection of $r_{s}$ on the plane  of the sky (i.e., $0 \leq R_s \leq
r_{s}$).

As an illustration, we consider  a simple case in which all sub-haloes
have  an  original  mass  $M_{s,0}=10^{11.5}\msunh$  at  the  time  of
accretion.    We  model   the  tidal   mass  loss   as   described  in
Section~\ref{sec:densprof}   and   adjust   their   density   profiles
accordingly.  Fig~\ref{fig:s11.5}  shows the full  ESD (including both
the host and the sub-halo  terms) around satellite galaxies located at
different halo-centric  distances in host haloes of  different masses. 
For simplicity  we assume  that $R_s =  r_{s}$, i.e., that  the radius
vector $\vec{r}_{s}$ is perpendicular to the line-of-sight.  Comparing
Figs~\ref{fig:s11.5} and~\ref{fig:s0}, one  sees that the contribution
of the sub-halo to the ESD  dominates at small $R$, as expected.  As a
comparison,   the  solid  curves   in  Fig~\ref{fig:s11.5}   show  the
$\Delta\Sigma_{s,{\rm   sub}}(R)$   of    sub-haloes   with   a   mass
$M_s=10^{11.5}\msunh$ (the contribution  from the host-term is ignored
here).   At   small  radii  these  are  slightly   higher  than  those
represented by  the dotted and  dashed curves, because  sub-haloes are
modeled to  have lost a fraction  $1-f_m$ of their  mass after having
been accreted  by the  host halo.  Note  that satellites  with smaller
$R_s$ have a somewhat lower ESD  at small radii. This owes to the fact
that the mass loss is  larger for sub-haloes at a smaller halo-centric
distance.

\section{Test using galaxies and groups in mock catalogues}
\label{sec_Nbody}

The analysis presented above shows that it is in principle possible to
use galaxy-galaxy lensing to probe the mass distribution of haloes and
sub-haloes. Unfortunately, we do not know a priori whether a galaxy is
a  central   galaxy  or  a  satellite  galaxy.    Since  the  expected
galaxy-galaxy lensing signal of  individual galaxies is very weak, one
has to combine the shear measured around a large number of galaxies to
obtain a  statistically significant  detection. The ESD  inferred from
such a measurement is therefore  the average over all lensing galaxies
used. It is  clear from the above, that if  that average combines both
central and  satellite galaxies, the  resulting $\Delta\Sigma(R)$ will
be  difficult  to  interpret,  especially  when  the  satellites  have
different $R_s$. 

However, if we could identify central and satellite galaxies a priori,
and  if we could  also determine  $R_s$ for  each satellite,  we could
measure  the  tangential  shear  of  central  and  satellite  galaxies
separately, thus constraining  the mean host halo mass  as well as the
mean sub-halo mass of the satellites directly.  Such an identification
of central  and satellite galaxies  requires a method to  decide which
galaxies belong to the same dark matter halo.  In a recent paper, Yang
\etal  (2005a)  developed a  new,  halo-based  group  finder which  is
particularly  successful in  grouping  galaxies in  a redshift  survey
according to  their common  dark matter haloes.   In this  section, we
apply  this  group finder  to  mock  redshift  catalogues and  examine
whether  the  membership  information   thus  obtained  is  useful  in
galaxy-galaxy lensing studies.  In  particular, we examine whether the
identification of central and  satellite galaxies inferred from such a
group catalogue  is sufficiently reliable  that it allows  an accurate
measurement of the masses of the corresponding host- and sub-haloes.
\begin{figure*}
  \centerline{\psfig{figure=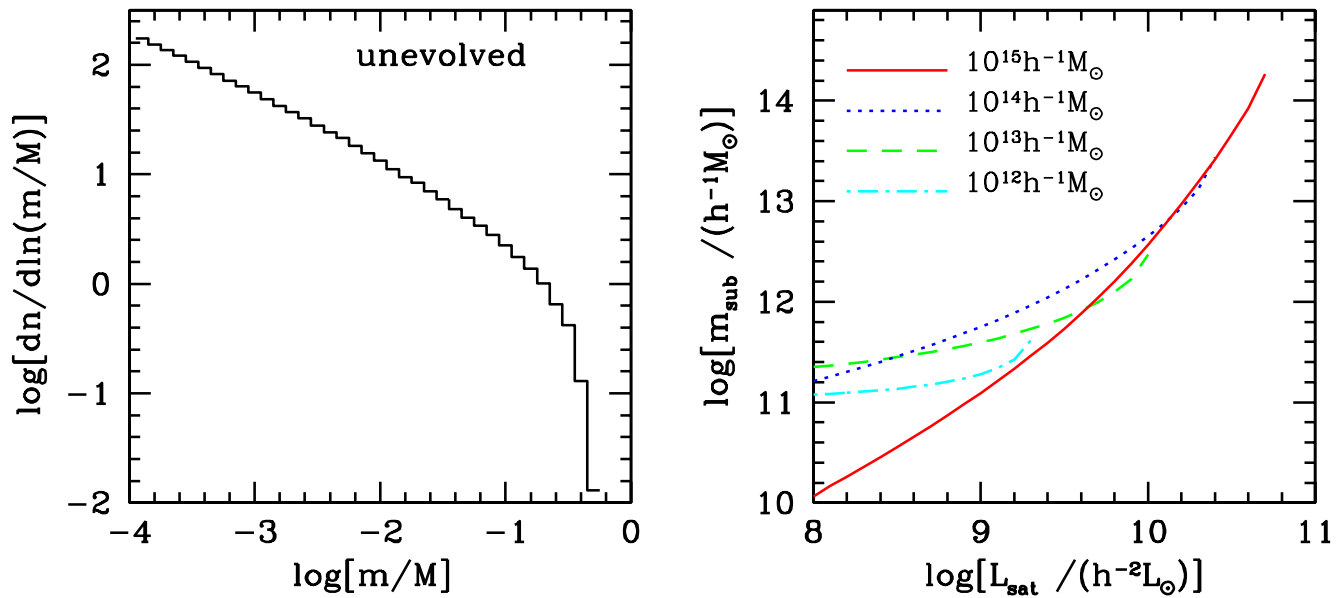,width=\hdsize}}
\caption{The left-hand panel shows the un-evolved mass function of dark 
  matter sub-haloes from van den Bosch \etal (2005a). Here `un-evolved'
  means that the masses correspond  to the masses of the sub-haloes at
  the time that they were accreted  (i.e., before any mass loss due to
  tidal stripping occurred).  As shown in van den Bosch \etal (2005a),
  this mass function is independent of  the mass of the host halo. The
  right-hand  panel shows  the relation  between the  luminosity  of a
  satellite  galaxy and  the mass  of the  sub-halo at  the  moment of
  accretion. This relation is obtained by comparing the number density
  of  satellite  galaxies  predicted  by  the  conditional  luminosity
  function  with the  number density  of sub-haloes  predicted  by the
  un-evolved sub-halo mass function shown in the left-hand panel.}
\label{fig:ms_ls}
\end{figure*}

\subsection{N-body simulations and  mock catalogues}
\label{sec:Nbody}

We use the results of a high-resolution N-body simulation to construct
mock galaxy catalogues. The simulation  was carried out on the VPP5000
Fujitsu  supercomputer  of the  National  Astronomical Observatory  of
Japan with  the vectorized-parallel P$^3$M  code (Jing \& Suto  2002). 
It  evolves  the distribution  of  the  dark  matter from  an  initial
redshift  of $z=72$  down  to $z=0$  in  a $\Lambda$CDM  `concordance'
cosmology  with  $\Omega_m=0.3$,  $\Omega_{\Lambda}=0.7$,  $h=H_0/(100
\kmsmpc)=0.7$ and  with a scale-invariant initial  power spectrum with
normalization  $\sigma_8=0.9$. The simulation  uses $512^3$  cold dark
matter  particles in a  periodic cube  of $100  \times 100  \times 100
h^{-3}  \Mpc^3$.  The  particle  mass is  equal  to $6.2\times  10^{8}
h^{-1}M_{\odot}$,  and dark  matter  haloes are  identified using  the
standard FOF algorithm  with a linking length of  $0.2$ times the mean
inter-particle separation.

In order to  construct realistic mock galaxy samples,  we populate the
haloes  with  galaxies,  using  the  conditional  luminosity  function
(hereafter CLF; Yang,  Mo \& van den Bosch 2003b;  van den Bosch, Yang
\&  Mo 2003), $\Phi(L  \vert M)$,  which gives  the average  number of
galaxies of luminosity $L$ that reside in a halo of mass $M$.  The CLF
model parameters used to construct the mock galaxy catalogue are given
in Table~1 of  van den Bosch \etal  (2005c) as ID \# 6.   We refer the
reader to Yang \etal (2004)  for details regarding the construction of
these  mock   galaxy  catalogues.    We  emphasize  though,   that  by
construction,  these  mock catalogues  match  the observed  luminosity
function  and   the  observed  clustering  strength   as  function  of
luminosity.

The CLF model  allows us to populate each halo  in the simulation with
galaxies of different luminosities.  We locate the brightest galaxy in
each halo at  the halo center and assume that  the other galaxies (the
satellites)   are   associated    with   dark   matter   sub-haloes.   
Unfortunately, we can  not use the actual sub-haloes  in the numerical
simulation itself, simply because  the resolution of the simulation is
not  sufficient to  resolve sub-haloes  in low  mass host  haloes.  In
addition, the survival and structure  of sub-haloes is affected by the
baryonic  component, which  is  not modeled  in our  dark-matter-only
simulation.  We  therefore follow a different  approach. We distribute
the satellite galaxies isotropically  throughout the host halo, with a
number density  distribution that  reproduces the dark  matter density
profile.  Note  that here  we use the  spherical NFW profile  with the
concentration obtained by Eke, Navarro \& Steinmetz (2001) in modeling
the  spatial distribution of  the satellite  galaxies relative  to the
halo center.   Next, we model the sub-halo  population associated with
these satellite galaxies adopting  a prescription similar to that used
in Kravtsov  \etal (2004), Vale  \& Ostriker (2006), and  Conroy \etal
(2006): we assume that the luminosity of the satellite has a monotonic
relation with  the sub-halo mass {\it  at the time of  its accretion}. 
In order  to establish such  a relation, we  use the mass  function of
sub-halo progenitors obtained by van den Bosch \etal (2005a), shown in
the  left-hand  panel  of  Fig.~\ref{fig:ms_ls}, and  the  conditional
luminosity function. The sub-halo mass function has been studied 
recently both in numerical simulations and galaxy-galaxy lensing 
observations (e.g. De Lucia \etal 2004; Natarajan \& Springel 2004; 
Natarajan \etal 2006). The mass-luminosity relations thus obtained are
shown  in  the  right-hand  panel of  Fig.~\ref{fig:ms_ls},  for  four
different host halo masses.  Using these luminosity-mass relations and
the     model      of     sub-halo     structure      described     in
Section~\ref{sec:densprof}, we  assign each satellite  a sub-halo mass
depending on  (i) its luminosity, (ii)  its host halo  mass, and (iii)
its distance from the center of the host halo.

Next we construct a mock redshift  catalogue. We put the center of the
simulation box  at a  comoving distance of  $200\mpch$ from  a virtual
observer.  Each galaxy  is given a redshift according  to its distance
from this  observer and its peculiar velocity  along the corresponding
line of sight.   We construct a flux-limited sample  by including only
those galaxies  that have an  apparent magnitude $m_{\rm  b_J}<19.30$. 
Finally,  we apply  our halo-based  group finder  to this  mock galaxy
catalogue  in  order  to  construct  a catalogue  of  galaxy  groups.  
Following  Yang \etal (2005b),  we assign  a halo  mass to  each group
according to its ranking in  total group luminosity and using the halo
mass function for the  standard $\Lambda$CDM cosmology (see Yang \etal
2005b for  more details).   Tests in Yang  \etal (2005b)  and Weinmann
\etal  (2006) have  shown  that,  on average,  the  group masses  thus
obtained are in good agreement with the input halo masses.

\begin{figure*}
\centerline{\psfig{figure=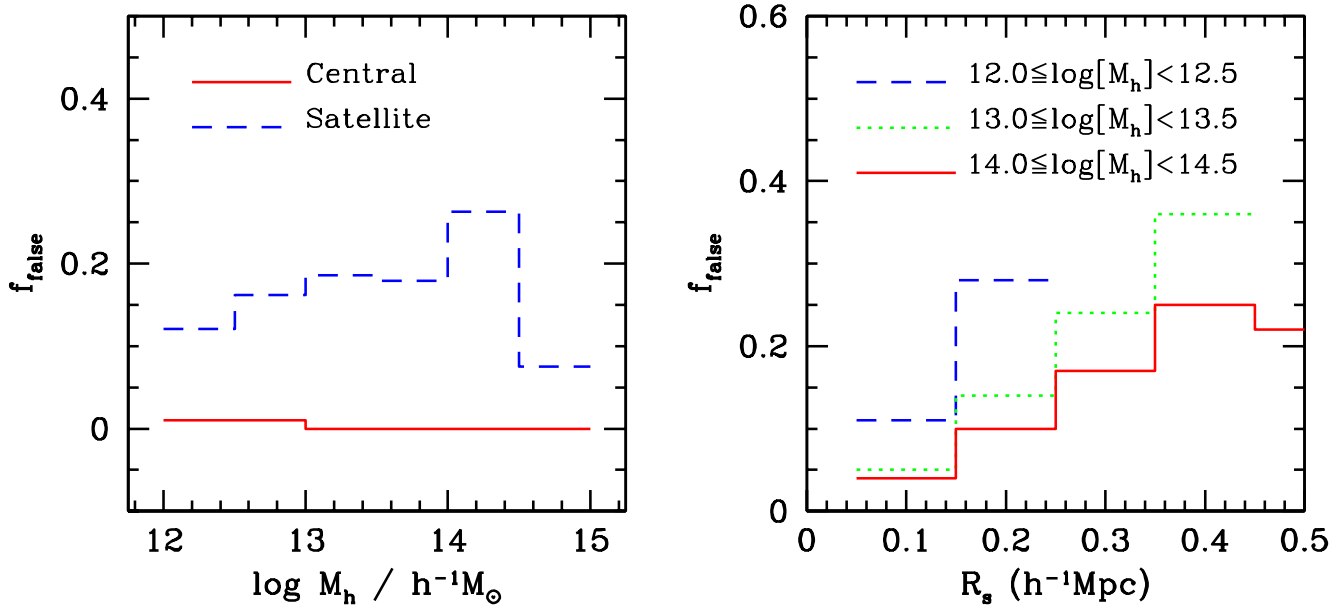,width=\hdsize}}
\caption{The left-hand panel shows the fractions of the false central 
  (solid line) and  satellite galaxies (dashed line) as  a function of
  group masses. The right-hand panel  shows the fractions of the false
  satellite galaxies  (interlopers) in  groups of different  mass bins
  (as  indicated)  as  a   function  of  the  projected  group-centric
  distances. }
\label{fig:false}
\end{figure*}

In what follows we refer to  the brightest galaxy in each group as the
central galaxy, and  to all other group members  as satellites. We use
the mock galaxy group catalogue to test how well galaxy-galaxy lensing
measurements around  these `central' and `satellite'  galaxies allow a
recovery  of their  host  and sub-halo  masses  respectively.  Due  to
interlopers (group members that do  not belong to the same dark matter
halo) and the fact that the  group finder may occasionally miss a halo
member, central and satellite galaxies  in the group catalogue are not
necessarily also central and satellite  galaxies in their real haloes. 
If this  confusion is  too large,  it will not  be possible  to obtain
reliable  estimates  of host  and/or  sub-halo  masses.  Clearly,  the
accuracy of  such an approach therefore  needs to be  tested, which is
the purpose of  the mocks constructed here. In  the left-hand panel of
Fig.~\ref{fig:false},  we  show the  fractions  of  the false  central
galaxies and  false satellite galaxies (interlopers) as  a function of
group  mass. The  fraction  of false  central  galaxies is  completely
negligible in groups of all  masses, while that of false satellites is
typically  well  below  20\%.   As  we have  pointed  out  in  section
~\ref{sec:satellite},  the  ESDs   around  satellite  galaxies  depend
strongly  on the  projected halo-centric  distances.  Therefore  it is
also important to check how  the interloper fractions of the satellite
galaxies depend on the projected group-centric distances.  The results
are shown  in the right-hand  panel of Fig.~\ref{fig:false}.   One may
notice that at small  projected group-centric distances the interloper
fractions are rather small, and  then increase (due to the larger area
covered) as  the projected group-centric distances  increase.  We will
discuss  later how  and to  what extent  this interloper  fraction may
affect our  measurements of the  ESDs and the extracted  properties of
the subhaloes.

\subsection{Lensing by galaxies in groups}
\label{sec_GMCCF}
\begin{figure*}
  \centerline{\psfig{figure=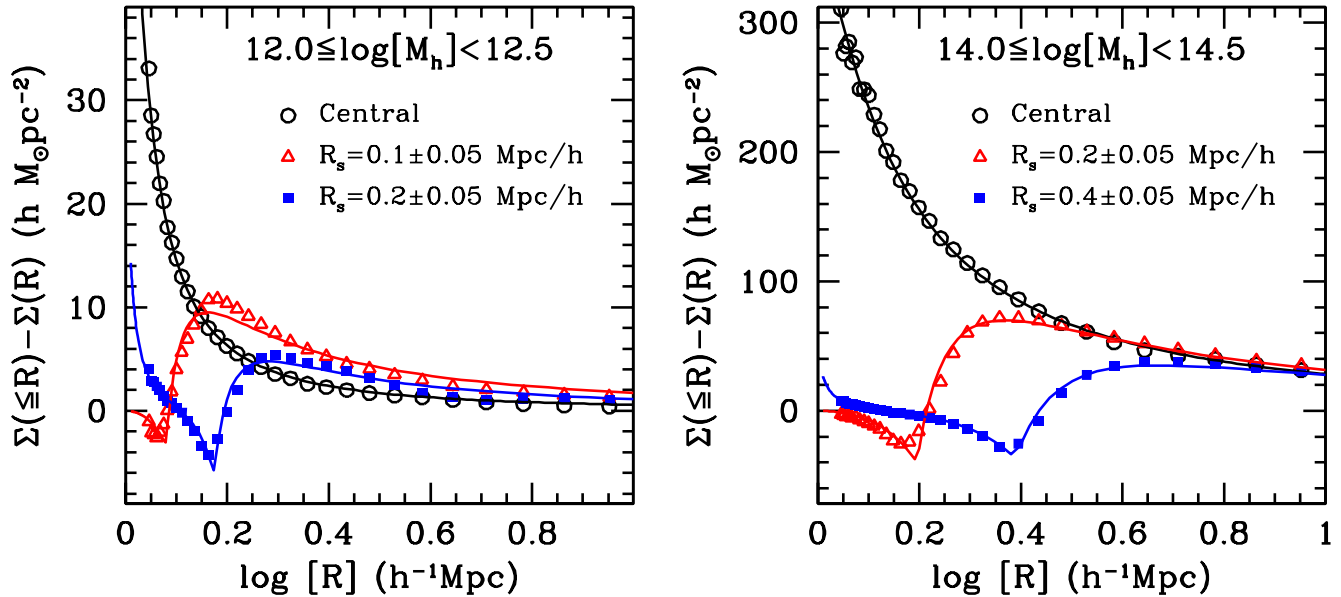,width=\hdsize}}
\caption{The ESDs measured  from the N-body  simulation  around  
  central  galaxies (open  circles) and  around satellite  galaxies at
  different  $R_s$, as  indicated (open  triangles and  solid squares)
  Note that  here satellite galaxies  are modeled without  dark matter
  sub-haloes. The two panels  correspond to different bins of assigned
  group mass, as  indicated with masses in $h^{-1}  \Msun$.  The lines
  in each panel are the best-fit models to the ESDs shown: For central
  galaxies, this model  has two free parameter ($M_h$  and $c$), while
  in  the case  of  satellite  galaxies it  has  four free  parameters
  ($M_h$,  $c$, $M_s$  and $R_s$).   Since  we do  not have  realistic
  `measurement' errors, we give each  `data point' equal weight in the
  fitting. See text for more details.}
\label{fig:data_no}
\end{figure*}

\subsubsection{The lensing signal around central galaxies}
\label{sec:smdc_cen}

We  first  consider  the  lensing  signal  around  the  central  group
galaxies.  In order to ``measure'' the ESD we project the positions of
galaxies and dark  matter particles onto a plane  perpendicular to the
line  of sight,  and estimate  the  mean dark  matter surface  density
contrast within  rings of different radii around  the galaxies.  Since
by definition  the background surface  density is subtracted,  the ESD
thus derived is independent of the depth of the projection (here equal
to the simulation box size), as long as this depth is much larger than
the dark matter correlation length.

The open  circles in Fig~\ref{fig:data_no}  show the $\Delta\Sigma(R)$
thus  obtained around  the central  galaxies in  groups  with assigned
masses in the  range $12.0 \le \log (M_h/\msunh)  < 12.5$ (left panel)
and $14.0 \le  \log (M_h/\msunh) < 14.5$ (right  panel), respectively. 
These ESDs  are monotonic  functions of radius,  and the  amplitude is
higher for the more massive  haloes, in agreement with the predictions
shown in  Section~\ref{sec_lensing}.  In Section~\ref{sec:extract}, we
investigate  the  accuracy  with  which  these  measurements  allow  a
recovery of the  mean host halo mass of these  galaxies. Note that the
group masses have  been assigned based on the  total group luminosity;
the lensing signal will provide a direct test of this mass.

\subsubsection{The lensing signal around satellite galaxies}
\label{sec:smdc_sat}

As discussed  in Section~\ref{sec_lensing}, the  galaxy-galaxy lensing
signal from satellite  galaxies depends not only on  the luminosity of
the satellite,  but also on  the properties of  the host halo  and the
halo-centric distance.  This means  that the tangential shear measured
around a large sample of  satellite galaxies in a given luminosity bin
is  very difficult  to  interpret.   Rather, one  can  use the  actual
information from the group  catalogue to sort the satellites according
to both  the mass of their  host halo (i.e., the  assigned group mass)
and the projected distance from the center of the group. Such a signal
is much easier to interpret, as it tightly constrains the contribution
of the $\Delta\Sigma_{s,{\rm host}}$-term to the shear measurements.

To illustrate the potential power  of this approach we measure the ESD
around  satellite  galaxies  in  our  mock group  catalog,  using  the
projected    distribution   of   dark    matter   particles    as   in
Section~\ref{sec:smdc_cen} above. Since the simulations do not resolve
the majority  of sub-haloes, and  since we did not associate satellite
galaxies in the  mock with the resolved sub-haloes,  this ESD reflects
the lensing signal that one would obtain if satellite galaxies are not
surrounded by dark matter sub-haloes.

The open triangles and solid squares in Fig~\ref{fig:data_no} show the
$\Delta\Sigma(R)$ around  satellite galaxies with  different projected
distances  from  their  central   group  galaxies,  as  indicated.  As
demonstrated in Section~\ref{sec:satellite}, the ESDs around satellite
galaxies are expected to become  similar to those around their central
galaxies at radii $R  \ga 3R_s$. However, in Fig~\ref{fig:data_no} the
ESDs  of the  satellite galaxies  have slightly  higher  amplitudes at
large radii  than those of  the corresponding central  galaxies.  This
systematic offset is due to the fact that we have combined groups in a
finite  mass range.  Since  more massive  haloes  (groups) in  general
contain a larger number  of satellite galaxies, any satellite-averaged
mean, such as the ESD, will be biased towards the more massive haloes.
A similar bias occurs when one tries to estimate the average halo mass
of  a stack of  host galaxies  from the  velocity dispersion  of their
satellite galaxies (see van den Bosch \etal 2004).

The ESD around satellite galaxies in groups with $10^{12} h^{-1} \Msun
\leq M_h < 10^{12.5} h^{-1} \Msun$  and with $R_s = 0.2\pm 0.05 h^{-1}
\Mpc$ (solid squares in left-hand  panel) reveals a small upturn at $R
\la  0.1  h^{-1}  \Mpc$.   However,  according  to  Fig.~\ref{fig:s0},
$\Delta\Sigma(R)$ should  go to zero at small  $R$.  This disagreement
with the  theoretical predictions is caused  by the fact  that a small
number  of  these  satellite  galaxies  in  the  group  catalogue  are
interlopers,  which in reality  are mainly  central galaxies  in lower
mass   haloes.    Since  the   interloper   fraction  increases   with
halo-centric   distance   (shown    in   the   right-hand   panel   of
Fig.~\ref{fig:false}), this explains why  this peak near $R=0$ is more
pronounced  for  satellites  with  a  larger $R_s$.   Clearly,  it  is
important  that any  group finder  used for  identifying  centrals and
satellites is properly calibrated to yield sufficiently low interloper
fractions.  As described in Yang  \etal (2005a), the group finder used
here has  been calibrated accordingly.  In  particular, the interloper
fraction is  constant (at $\sim  20$ percent) with group  mass, unlike
the more standard friends-of-friends  method which typically yields an
interloper  fraction  that  increases systematically  with  decreasing
group mass (see Fig.~7 in Yang \etal 2005a).
\begin{figure*}
\centerline{\psfig{figure=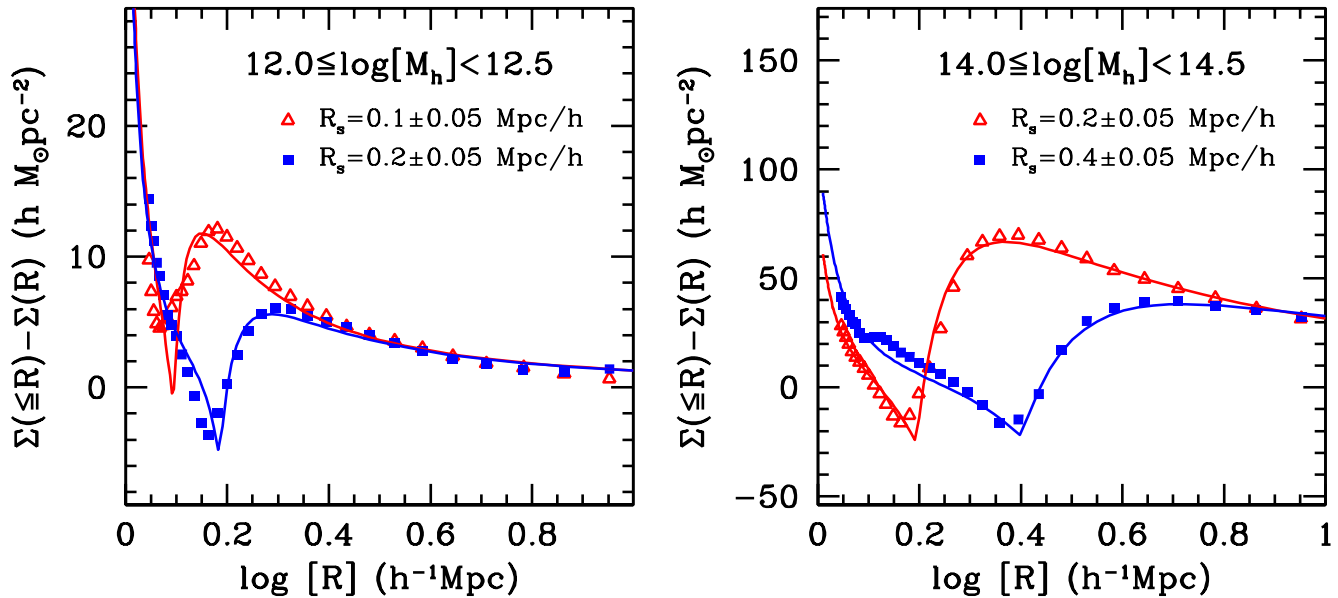,width=\hdsize}}
\caption{Same as Fig.~\ref{fig:data_no}, except that here satellite
  galaxies are associated with dark matter sub-haloes, as described in
  the text. Note that this enhances the ESD at small $R$.}
\label{fig:data_sub}
\end{figure*}

We  now add  the  contribution  of the  dark  matter sub-haloes.   The
density distribution of a sub-halo surrounding a satellite galaxy of a
given luminosity, located  at a given halo-centric distance  in a host
halo    of   a   given    mass   is    modeled   as    described   in
Section~\ref{sec:Nbody}, and its contribution to the lensing signal is
computed  by  integrating $\rho_s(r)$  along  the line-of-sight.   The
resulting ESDs  are shown in Fig.~\ref{fig:data_sub}. As  we have seen
before, the  contribution of the  dark matter sub-haloes  enhances the
ESD around satellite galaxies on small scales.

\subsection{Estimating the masses of haloes and sub-haloes from
galaxy-galaxy lensing}
\label{sec:extract}

Having  estimated  the lensing  signal  that  one  could in  principle
measure  around  central   and  satellite  galaxies  (given  virtually
infinite   signal-to-noise  in   the  shear   measurements),   we  now
investigate to what extent such a signal allows to recover the average
masses and  concentration parameters of  both the host haloes  and the
sub-haloes.

\subsubsection{The masses of haloes}
\label{sec:ext_host}

\begin{figure*}
\centerline{\psfig{figure=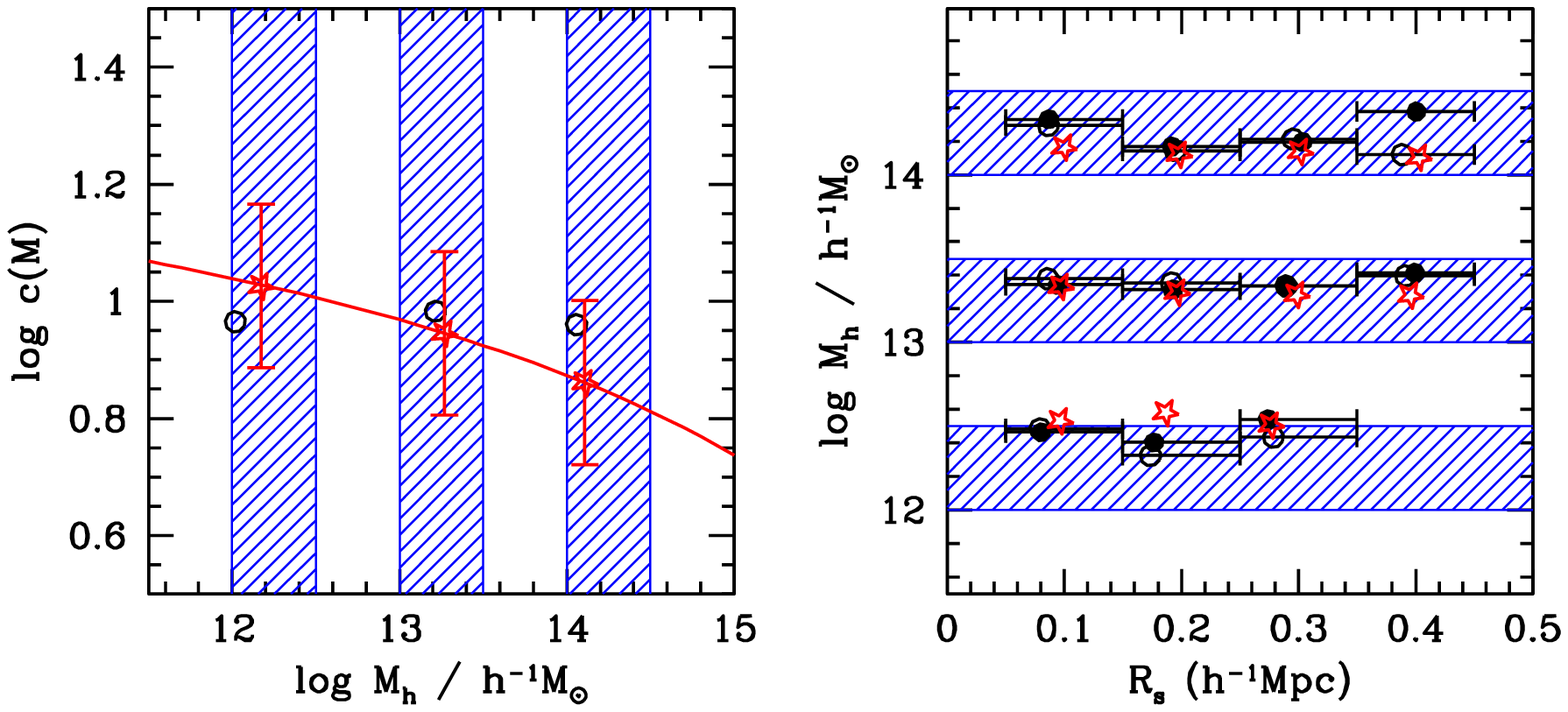,width=\hdsize}}
\caption{Properties of host haloes recovered from fitting the ESDs
  around  central galaxies  (left-hand panel)  and  satellite galaxies
  (right-hand panel)  in the mock group catalogue.   The hatched areas
  in both panels  indicate the range of assigned  masses of the groups
  from  which the central  and satellite  galaxies are  selected.  The
  open  circles  in  the  left-hand  panel  indicate  the  masses  and
  concentrations  of the  NFW haloes  that  best-fit the  ESDs of  the
  central group galaxies.  The asterisks show the true  mean host halo
  masses and the corresponding  halo concentrations obtained using the
  model of  Eke \etal (2001; shown  as solid line).   The errorbars on
  the asterisks  correspond to $\Delta \log c=0.14$ which reflects the
  typical scatter in $c$ at a given halo mass. The solid dots and open
  circles  in the  right-hand panel  indicate the  best-fit  host halo
  masses and group-centric radii obtained from fitting the ESDs around
  satellite   galaxies  with   and  without   sub-halo  contributions,
  respectively.  The  horizontal error  bars on these  points indicate
  the ranges of group-centric radius,  $R_s$, that were used to select
  the  satellite  galaxies from  the  group  catalogue. The  asterisks
  indicate the true mean host halo masses and the true mean $R_s$, for
  the corresponding satellite galaxies.}
\label{fig:mh}
\end{figure*}

We start by testing how  accurately the lensing signal around galaxies
classified as `centrals' by our  group-finder allows a recovery of the
mean mass  and concentration of their  host haloes. To  that extent we
use the  assigned group masses  as a `pre-selection', and  compute the
$\Delta\Sigma_c(R)$ around central  galaxies in three group-mass bins,
two  of which are  shown in  Fig~\ref{fig:data_no}. Assuming  that the
density distributions of dark matter  haloes are well described by NFW
profiles,      we     fit     these      $\Delta\Sigma_c(R)$     using
eq.~(\ref{eq:SMDC_cen}) with  $c$ and  $M_h$ as free  parameters.  The
best  fits are  shown as  solid lines  in  Fig.~\ref{fig:data_no}, and
should be compared to the open circles.

The corresponding  best fit values of  $M_h$ and $c$ are  shown in the
left panel  of Fig~\ref{fig:mh} as the open  circles.  For comparison,
the  asterisks show  the  true mean  halo  masses that  host the  mock
galaxies used to determine $\Delta\Sigma_c(R)$, while the hatched area
indicates the  bin of assigned  group masses.  Note that  the best-fit
$M_h$ is  in good agreement with  the true average, and  that both lie
well  within  the  range  of  assigned group  masses.   The  recovered
concentration parameters  also match the expected  values predicted by
Eke,  Navarro  \&  Steinmetz  (2001),  indicated by  the  solid  line,
reasonably  well.   Although there  are  slight  deviations, they  are
significantly smaller than the typical  scatter in $c$ for haloes of a
given  mass, $\Delta  \log c  \sim 0.14$,  indicated by  the errorbars
attached  to  the  asterisks  (e.g.  Jing 2000,  Bullock  \etal  2001,
Wechsler \etal 2002).

In the case of real galaxy-galaxy lensing data, the errors on
$\Delta\Sigma_c(R)$ will depend on the details of the selection of the
lens-source samples and on the quality of the observational data.  We do not
model such errors here; instead we give each data point an equal weight in the
fitting. We have also tested the impact of changing the weight scheme, the
change in the best fitting parameters are very small. Note that the errorbars
from recent galaxy-galaxy lensing measurements show different radial
dependences (e.g.  Mckay \etal 2001; Mandelbaum \etal 2005a), the absence of
realistic errorbars on our `measurements' prevents us from putting meaningful
confidence levels on the best-fit parameters.

The ESDs  measured around satellite galaxies  also contain information
regarding the  mass of the host halo.  In order to see  how well these
lensing measurements  allow to  recover the masses  of their  host and
sub-haloes  we fit  with  eq.~(\ref{delsigsat}) the  $\Delta\Sigma_s(R
\vert  R_s)$  for  a number  of  bins  in  $R_s$.  We have  four  free
parameters in the fit: $M_h$, $c$, $M_s$ and $R_s$ (subscripts $h$ and
$s$ refer to the host and sub-haloes, respectively).  Note that $c$ is
the concentration of  the host halo, not that  of the sub-halo.  Since
the inner part of the ESDs  cannot be measured with high precision, we
assume that  all sub-haloes have a  concentration $c_s =  10$. We will
discuss later  in section  ~\ref{sec:ext_sub} how this  assumption may
affect the extracted properties  (i.e. masses) of the sub-haloes. Note
also  that  we let  $R_s$  be a  free  parameter,  eventhough we  only
consider satellite  galaxies in a  relatively narrow bin in  $R_s$ (as
determined from the group catalogue).

The   best-fit   curves   are    given   by   the   solid   lines   in
Figs.~\ref{fig:data_no}   and~\ref{fig:data_sub}.   Note   that  these
best-fit models are  in reasonable agreement with the  `measured' ESD. 
The slight deviations are  due to interlopers (i.e., errors associated
with the  group catalogue)  and to  the fact that  the model  does not
account for the fact that the  data corresponds to a range in host and
sub-halo masses (i.e., $\Delta\Sigma_s(R \vert R_s)$ reflects a linear
combination of  many slightly different ESDs). The  best-fit host halo
masses are shown  in the right panel of  Fig.~\ref{fig:mh}: solid dots
are the results obtained using the mock in which we do not include the
analytical sub-haloes (Fig.~\ref{fig:data_no}), while the open circles
show   the    results   obtained   when    sub-haloes   are   included
(Fig.~\ref{fig:data_sub}).  The agreement with the true mean host halo
masses, indicated by the asterisks, is extremely good, suggesting that
the lensing signal from satellite galaxies basically yields an equally
good  measurement of  the  host halo  mass  as that  from the  central
galaxies.  Thus, with a  well-defined group catalog, the properties of
the host haloes  can be accurately recovered from  the ESDs using both
central and satellite galaxies.

There  is one small  caveat though.   Note that  the host  halo masses
obtained  from the  central  galaxies are  somewhat  lower than  those
obtained from the satellites.  This is  again due to the fact that the
ESDs of  satellite galaxies  are weighted by  the number  of satellite
galaxies;  since there  are more  satellite galaxies  in  more massive
groups, more massive systems receive a larger weight in the averaging.
In particular, groups that consist only of a single member (which then
by   definition   is   a   central   galaxy),   only   contribute   to
$\Delta\Sigma_c$, but not to  $\Delta\Sigma_s$.  Since such a group is
likely to have  a relatively low mass compared to  the other groups in
the same mass bin, this will bias the value of $M_h$ inferred from the
satellites high with  respect to the value of  $M_h$ inferred from the
central galaxies.  Note, however, that this does not reflect an error;
the average true  halo mass of systems with at  least one satellite is
simply  larger  than  the  average  true  halo  mass  of  all  systems
(including those with zero satellites).  This is also evident from the
fact  that  the  asterisks  in  the  left and  right  hand  panels  of
Fig.~\ref{fig:mh} indicate different masses.  Thus, if the host masses
inferred from $\Delta\Sigma_c$ and  $\Delta\Sigma_s$ do not agree with
each other,  this does not  necessarily indicate an  inconsistency. It
may also simply reflect a  `selection effect', in that the ESDs around
central and  satellite galaxies are contributed  by somewhat different
host  halo populations.  As a simple test, we weight the ESDs for 
central galaxies with the number of satellite galaxies at different
group-centric distances. The recovered host halo masses are now in much
better consistent with those recovered from the ESDs around satellite 
galaxies. 

\subsubsection{The masses of sub-haloes}
\label{sec:ext_sub}

\begin{figure*}
\centerline{\psfig{figure=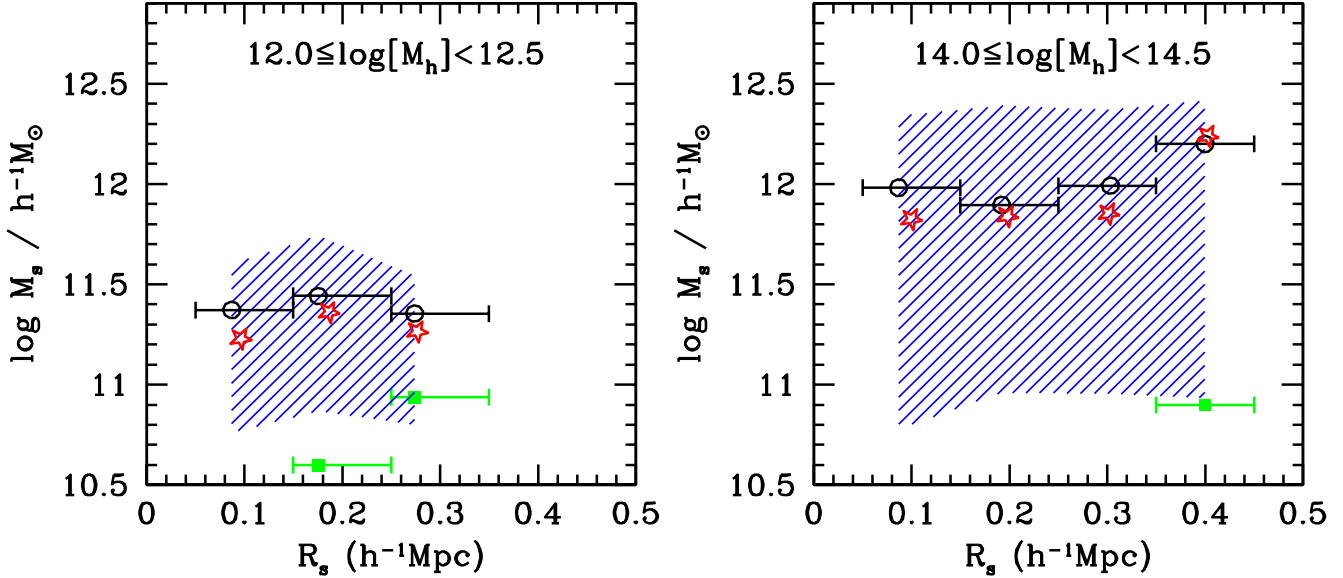,width=\hdsize}}
\caption{Open circles indicate the best-fit sub-halo masses and 
  group-centric radii obtained from  fitting the ESDs around satellite
  galaxies in the mock group  catalogue in which each satellite galaxy
  has  been   assigned  a  dark   matter  sub-halo  as   described  in
  Section~\ref{sec_Nbody}.  Different  panels correspond to  groups in
  different bins of assigned  mass, as indicated, while the horizontal
  error bars indicate the ranges of $R_s$ that were used to select the
  satellite galaxies from the group catalogue.  The hatched areas show
  the 90  percentile ranges of  true (retained) sub-halo  masses.  The
  asterisks indicate the  true mean sub-halo masses and  the true mean
  $R_s$, for the corresponding  satellite galaxies. Finally, the solid
  squares  (also  with  horizontal  errorbars) indicate  the  best-fit
  values of  $M_s$ and $R_s$ obtained  by fitting the  ESDs around the
  satellite galaxies in the mock {\it without} dark matter sub-haloes.
  The fact that  these best-fit values are not equal  to zero, as they
  should be,  owes to the  contamination due to interlopers.  See text
  for a detailed discussion.}
\label{fig:msub}
\end{figure*}
\begin{figure*}
\centerline{\psfig{figure=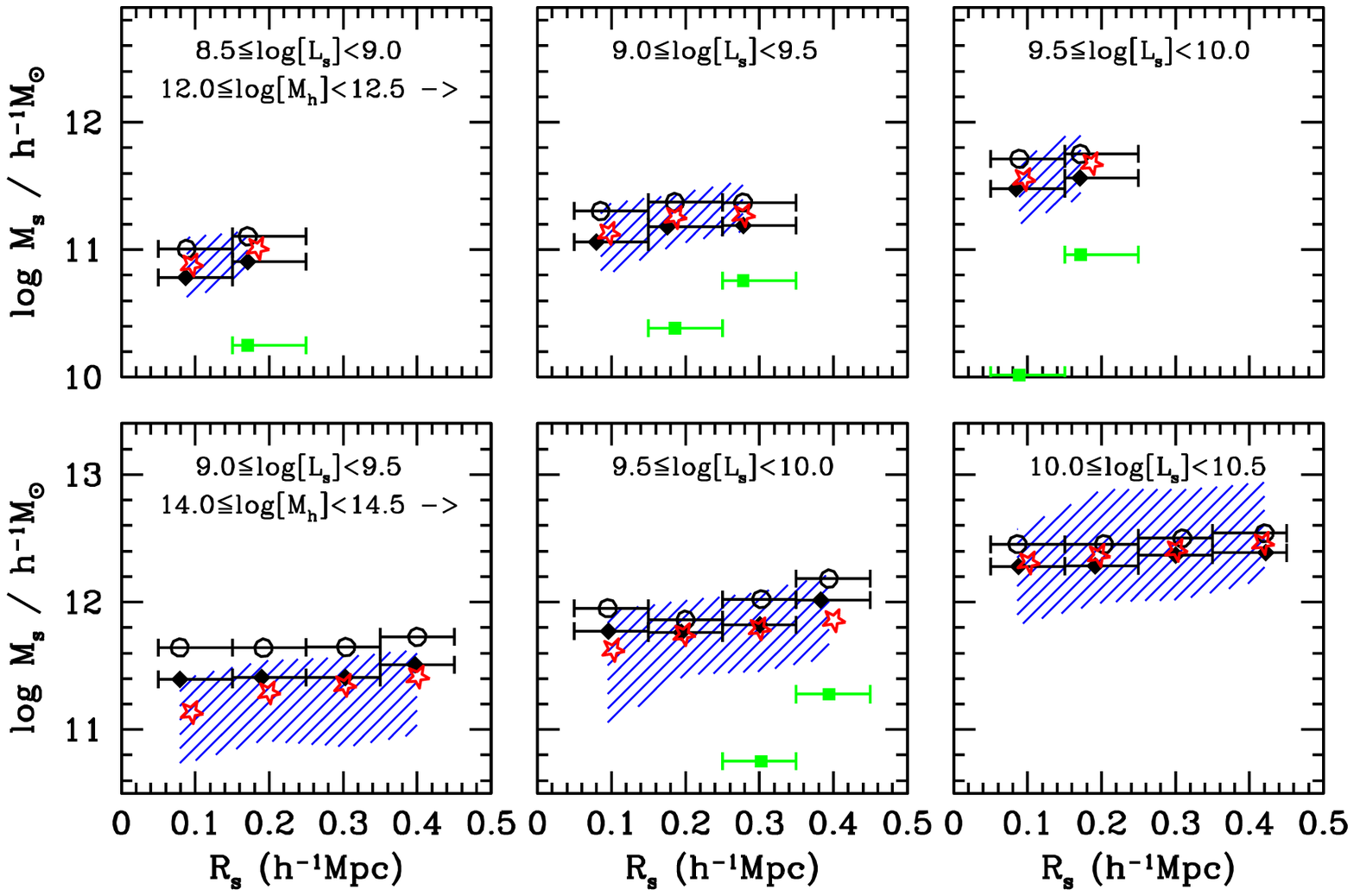,width=\hdsize}}
\caption{Same as Fig.~\ref{fig:msub}, except that here we show the
  results  for  different  bins  in  satellite  luminosity  $L_s$,  as
  indicated  in $h^{-2}  \Lsun$.  As  in Fig.~\ref{fig:msub}  the open
  circles show  the best-fit  sub-halo masses and  group-centric radii
  obtained  fitting the ESDs  assuming NFW  sub-haloes with  $c_s=10$. 
  The solid diamonds show the same quantities but obtained by assuming
  sub-haloes  modeled by  eq.~(\ref{eq:rhos}) with  mean  $\langle r_s
  \rangle \approx (R_s+r_{{\rm vir},h})/2$.}
\label{fig:msub_L}
\end{figure*}

The four-parameter fits  to the ESDs of satellite  galaxies also yield
best-fit  values for  the average  sub-halo  mass, $M_s$  and for  the
projected  distance, $R_s$.  The  open circles  in Fig.~\ref{fig:msub}
show the best-fit values of $M_s$  and $R_s$ as obtained from the mock
in which  we added analytical sub-haloes around  each satellite galaxy
(i.e., these correspond to the ESDs shown in Fig.~\ref{fig:data_sub}).
For  comparison,  the  asterisks  show  the  corresponding  true  mean
values\footnote{For  interlopers   that  are  centrals,   rather  than
  satellites,  we use  the host-halo  mass in  the computation  of the
  averages.}.  The  solid squares in  Fig.~\ref{fig:msub} indicate the
best-fit values of $M_s$ obtained from the mock in which no sub-haloes
are   included   (i.e.,   corresponding   to   the   ESDs   shown   in
Fig.~\ref{fig:data_no}). The  fact that these best-fit  values are not
equal to zero, as they should  be, owes to the presence of interlopers
that in  reality are centrals.  These best-fit  sub-halo masses should
therefore be considered a contamination of the measurement in the case
{\it with}  subhaloes.  Typically  this contamination is  much smaller
than the actual mean sub-halo  mass, especially at small $R_s$. In the
cases  examined here,  the largest  contamination we  find is  for the
satellites  with $R_s  =  0.3 \pm  0.05  h^{-1} \Mpc$  in groups  with
$10^{12}  h^{-1}\Msun \leq  M_h  < 10^{12.5}  h^{-1}\Msun$, where  the
best-fit $M_s$ inferred from the mock without sub-haloes is 38 percent
of that obtained with sub-haloes.  In all other cases this fraction is
much lower.

The  shaded areas  in  Fig.~\ref{fig:msub} outline  the 90  percentile
range of the real sub-halo  masses, which clearly is very broad.  This
owes  to the  fact that  we  have added  the lensing  signal from  all
satellite galaxies,  irrespective of  their luminosity. In  our model,
and most  likely also in  reality, more luminous satellites  reside in
more  massive  sub-haloes. This  suggests  that  the  actual range  of
sub-haloes probes  may be narrowed by selecting  satellite galaxies of
similar  luminosities.  Fig.~\ref{fig:msub_L}  shows  the ESDs  around
satellite  galaxies in  different  luminosity bins  and  in groups  of
different  (assigned) masses.  As  in Fig.~\ref{fig:msub},  the shaded
areas indicate the  90 percentile ranges of the  true sub-halo masses. 
As expected, these are now  much narrower. The remaining width owes to
the finite widths of the  ranges in satellite luminosity, $L_s$, group
mass, $M_h$,  and projected  separation, $R_s$, and  to the  fact that
projection causes systems with different $r_{s}$ to contribute to the
same $R_s$.  Note that in our model the sub-halo mass is a function of
$M_h$, $L_s$ and $r_{s}$.

The  open  circles in  Fig.  ~\ref{fig:msub_L}  indicate the  best-fit
sub-halo  masses, obtained by  fitting the  corresponding ESDs  with a
four  parameter model  as described  above.  Note  that  the recovered
sub-halo masses  are slightly larger than the  input value, especially
for fainter satellite galaxies in massive groups.  We propose that the 
main reason for
this  is  that  a  stripped  halo  has a  steeper  mass  profile  (see
eq.~\ref{eq:rhos}),  and the assumption  of a  fixed NFW  profile with
$c_s=10$ is not sufficiently accurate.
As a simple  test of this hypothesis, we consider a simple model in  
which we assume that
the  mean halo-centric  distances of  the satellite  galaxies  in each
sample is $\langle r_s \rangle\approx (R_s+r_{{\rm vir},h})/2$.  Then,
we use eq.~\ref{eq:rhos}  to model the mean profile  of the sub-haloes
in  consideration. This  model  ensures that  sub-haloes with  smaller
$R_s$, which  are expected to have  suffered more mass  loss, are more
concentrated. The  best-fit sub-halo  masses obtained from  this model
are shown  in Fig.~\ref{fig:msub_L} as  the solid diamonds. As  can be
seen, the best-fit masses are  systematically smaller than in the case
of $c_s=10$, and are in better agreement with the true averages.

In  conclusion,  with  a  well-defined  group  catalog,  the  ESDs  of
satellite galaxies  obtained from galaxy-galaxy  lensing, with signals
properly  stacked according  to the  group masses  (inferred  from the
total  group  luminosities)  and  the luminosities  and  group-centric
distances of  satellite galaxies, can be  used to probe  the masses of
sub-haloes associated with satellite galaxies. 

\subsubsection{What if the brightest galaxy is off-center?}
\label{sec:BCG}

In the analysis presented above,  the brightest halo galaxy is assumed
to  reside  at  the halo  center.  In  reality,  the location  of  the
brightest galaxy may  not be exactly at the halo  center, and this may
affect the predicted ESDs around the central and satellite galaxies.
Here we examine the importance of this effect.

In a  recent study  of the phase  space distribution of  the brightest
halo  galaxies in  galaxy groups,  van den  Bosch \etal  (2005b) found
that, for haloes with masses $M_h>10^{13}\msunh$, the deviation of the
location  of the  brightest galaxy  from the  halo center  is  about 3
percent of the  halo virial radius. Based on  SPH simulations, Berlind
(2003) also found that the `central'  galaxy in a dark matter halo may
deviate from the  position of the most bound particle  by 2 percent of
the  halo virial  radius. In  order  to quantify  how such  deviations
impact on the ESDs, we perform  the following test. We construct a new
mock catalogue in which we  assume that the position of each brightest
halo galaxy deviates from the most bound particle with an amount given
by a Gaussian distribution with a dispersion equal to 3 percent of the
halo virial radius.   Next we run our group finder  over this new mock
catalogue,  compute  the  ESDs   around  the  central  and  satellites
galaxies, and  fit them as  described above to determine  the best-fit
halo masses and concentrations. We  find that the off-centering of the
central  galaxies has  an almost  negligible effect,  except  that the
best-fit  concentration parameters  of host  haloes are  reduced  by a
modest 15 percent. Such a reduction arises from the fact 
that the off-centering decreases the ESD at small radius, similar 
to what happens to the ESD around satellite galaxies.
Thus, the ability to measure the host halo masses is  not compromised 
by the fact  that central galaxies may be (slightly) off-centered from 
the gravitational center of their dark matter halo.

\section{Discussion and conclusions}
\label{sec_discussion}

The  weak galaxy-galaxy  lensing  signals from  central and  satellite
galaxies (at different halo-centric distances)
  are very  different. So  far, this  difference has  not been
fully  exploited.   In fact,  in  all  previous  studies it  has  been
considered a nuisance,  rather than a source of  valuable information. 
Most  galaxy-galaxy lensing  analyses have  either simply  ignored the
differences between central and satellite galaxies, or have focused on
isolated   galaxies  in   an  attempt   to  minimize   the  disturbing
contribution of satellite galaxies. An alternative approach, pioneered
by Guzik  \& Seljak  (2002), is to  model the combined  lensing signal
from central and  satellites. However, no study to  date has attempted
to measure the lensing signals  of centrals and satellites at different
halo-centric distances separately. 
This  simply owes to  the fact  that it  is not  a priori  clear which
galaxy  is a central  galaxy and  which galaxy  a satellite.   This is
unfortunate, as  a separation  of these two  components would  allow a
much  cleaner  measurement of  the  halo  masses  (and their  detailed
density  profiles) hosting  central  galaxies. In  addition, it  would
allow for a direct measurement of the masses (and density profiles) of
sub-haloes hosting  satellite galaxies.  Furthermore,  since the shear
around satellite galaxies also  harbors information regarding the host
haloes in  which their  sub-haloes reside, one  can use  their lensing
signal to obtain an independent measurement of the host halo masses.

In order to  exploit this richness in information,  it is crucial that
one  has a  reliable technique  for separating  central  and satellite
galaxies. In this  paper, we have investigated to  what extent this is
feasible  with the  halo-based group  finder developed  by  Yang \etal
(2005a).   This group  finder is  straightforward to  apply  to galaxy
redshift  surveys, and  has been  calibrated to  yield  low interloper
fractions.   The latter  is  important to  minimize  the confusion  of
centrals  and  satellites.   In  order  to test  our  methodology,  we
constructed  detailed mock  redshift surveys  from  N-body simulations
that  are populated  with  galaxies using  the conditional  luminosity
function.   Application of the  halo-based group  finder to  this mock
redshift survey yields  a large catalogue of mock  galaxy groups, from
which central galaxies are identified  as the brightest galaxy in each
group.  All other group members are considered to be satellites.

The  group catalogue  is not  only  useful to  separate centrals  from
satellites.   It  also  yields,  for  each  satellite,  the  projected
separation, $R_s$, to it  central galaxy. This is extremely important,
since the  contribution of  the host halo  to the  gravitational shear
around a  satellite galaxy depends  strongly on $R_s$.   Therefore, in
order  to   facilitate  a  meaningful  interpretation   of  the  shear
measurements around  satellite galaxies, it  is crucial that  one only
stacks the data from a  relatively narrow bin in $R_s$.  Another piece
of useful  information from the group  finder are the  masses that one
can assign  to each  group based  on the total  luminosity of  all its
members. This mass  can be used to `pre-select'  central and satellite
galaxies  in  relatively  narrow  bins  in  group  mass,  which  again
facilitates the interpretation of the lensing measurements.

To  test these  ideas we  computed the  excess surface  density (ESD),
which  is  the  observable  that  can be  obtained  from  the  lensing
measurements, around each central and satellite galaxy in the $N$-body
simulation.   We have demonstrated  that by  stacking the  ESDs around
central galaxies within different  group mass bins, the average masses
and concentrations  of their host  haloes can be  accurately recovered
from  the data.   In  addition, we  have  shown that  the ESDs  around
satellite galaxies,  when stacked according  to group-centric distance
and group  mass, allow an equally  accurate recovery of  the masses of
their corresponding sub-haloes, as well as that of their host haloes.

We therefore conclude that a combination of galaxy-galaxy lensing measurements
with a galaxy group catalogue extracted from a large redshift survey, such as
the Sloan Digital Sky Survey, in principle allows for accurate measurements of
the masses and concentrations of host haloes around central galaxies and
sub-haloes around satellite galaxies.  However, it is important to realize
that we have not attempted to mimic realistic observations. Rather, we have
simply assumed infinite accuracy in the measurements of the ESDs.  In reality,
resolution issues due to the finite sampling of the shear field and errors in
the shear measurements, may cause a substantial reduction of the accuracy with
which this methodology can be applied.  In a forthcoming paper, we will apply
this method to realistic mock shear maps taking into account various
observational effects to test its feasibility.

%%%%%%%%%%%%%%%
% Acknowledgments
%%%%%%%%%%%%%%%

\section*{Acknowledgment}

We thank  Michael Hudson for useful  comments. XY is  supported by the
{\it One Hundred  Talents} project of the Chinese  Academy of Sciences
and grants  from NSFC  (No.10533030). YPJ is  supported by  the grants
from  NSFC (Nos.10125314,  10373012, 10533030)  and from  Shanghai Key
Projects in Basic research (No.  04JC14079 and 05XD14019).

%%%%%%%%%%%%%%%
% Bibliography
%%%%%%%%%%%%%%%

\label{lastpage}


\begin{thebibliography}{}

\bibitem[]{BA04.1}
Bartelmann, M. \& Meneghetti, M. 2004, A\&A, 418, 413
  
\bibitem[]{Bar96} 
Bartelmann M., 1996, A\&A , 313, 697
  
\bibitem[]{Ben02} 
Benson A.J., Lacey C.G., Baugh C.M., Cole S., Frenk C.S., 2002,
\mnras, 333, 156
  
\bibitem[]{Bel02} 
Berlind A.A., Weinberg D.H., 2002, \apj , 575, 587
  
\bibitem[]{Bel03} 
Berlind A.A., \etal , 2003, ApJ, 593, 1

\bibitem{Brainerd96} 
Brainerd T.G, Blandford R.D., Smail I. 1996, \apj , 466, 623
  
\bibitem[]{Bul01} 
Bullock J.S., Kolatt T.S., Sigad Y., Somerville R.S., Kravtsov A.V., 
Klypin A.A., Primack J.R., Dekel A., 2001, \mnras , 321, 559
  
\bibitem[]{Cole00} 
Cole S., Lacey C.G., Baugh C.M., Frenk C.S., 2000, \mnras, 319, 168

\bibitem[]{Con05} 
Conroy C., et al., 2005, \apj, 635, 982
  
\bibitem[]{Con06} 
Conroy C., Wechsler R.H., Kravtsov A.V., 2006, \apj , 647, 201
  
\bibitem[]{Coo05} 
Cooray A., 2005, \mnras , 364, 303
  
\bibitem[]{Coo06} 
Cooray A., 2006, \mnras , 365, 842
  
\bibitem[]{Cro06} 
Croton D.J., et al., 2006, \mnras , 365, 11
 
\bibitem[]{Da03} 
Dahle H., Hannestad S., Sommer-Larsen J., 2003, \apj , 588, 73

\bibitem[]{del96}
dell'Antonio I.P., Tyson J.A., 1996, \apj, 473, L17

\bibitem[]{De04} 
De Lucia G., Kauffmann G., Springel V., White S.D.M., Lanzoni B., 
Stoehr F., Tormen G., Yoshida N., 2004, \mnras , 348, 333
  
\bibitem[]{Eke01} 
Eke V.R., Navarro J.F., Steinmetz M., 2001, \apj , 554, 114
  
\bibitem{Fischer00} 
Fischer P., et al. 2000, AJ, 120, 1198
  
\bibitem[]{Far01} 
Fardal M.A., Katz N., Gardner J.P., Hernquist L., Weinberg D.H., 
Dav'e R., 2001, \apj , 562, 605
  
\bibitem[]{Gao04} 
Gao L., White S.D.M., Jenkins A., Stoehr F., Springel V., 2004, 
\mnras, 355, 819

\bibitem[]{Gent04} 
Gentile G., Salucci P., Klein U., Vergani D., Kalberla P., 2004, MNRAS,
351, 903

\bibitem[]{Gent05} 
Gentile, G., Burkert, A., Salucci, P., Klein, U., \& Walter, F.\ 
2005, \apj, 634L, 145

\bibitem[]{Gri96}
Griffiths R.E., Casertano S., Im M., Ratnatunga K.U., 1996, \mnras,
282, 1159
  
\bibitem[]{Guz01} 
Guzik J., Seljak U., 2001, \mnras , 321, 439
  
\bibitem[]{Guz02} 
Guzik J., Seljak U., 2002, \mnras , 335, 311
  
\bibitem[]{Hay03} 
Hayashi E., Navarro J.F., Taylor J.E., Stadel J., Quinn T., 2003, 
\apj , 584, 541
  
\bibitem{Hoekstra03} 
Hoekstra H., Franx M., Kuijken K., Carlberg R.G., Yee H.K.C., 2003, 
\mnras, 340, 609
  
\bibitem{Hoekstra04} 
Hoekstra H., Yee H.K.C., Gladders M.D., 2004, \apj , 606, 67
  
\bibitem{Hoekstra05} 
Hoekstra H., Hsieh B.C., Yee H.K.C., Lin H., Gladders M.D., 2005, 
\apj, 635, 73
  
\bibitem{Hudson98} 
Hudson M.J., Gwyn S.D.J., Dahle H., Kaiser, N. 1998, \apj, 503, 531
  
\bibitem[]{JMB98} 
Jing Y.P., Mo H.J., B\"{o}rner G., 1998, \apj , 494, 1

\bibitem[]{JS00} 
Jing Y.P., 2000, \apj , 535, 30 

\bibitem[]{JS00a} 
Jing Y.P., Suto Y., 2000, \apj , 529L, 69 
  
\bibitem[]{JS02} 
Jing Y.P., Suto Y., 2002, \apj , 574, 538 
  
\bibitem[]{Kau93} 
Kauffmann G., White S.D.M., Guiderdoni B., 1993, \mnras, 264, 201
  
\bibitem[]{Kauff04} 
Kauffmann G., White S.D.M., Heckman T.M., Menard B., Brinchmann J., 
Charlot S., Tremonti C., Brinkmann J. 2004, \mnras , 353, 713
  
\bibitem[]{Katz96} 
Katz N., Weinberg D.H., Hernquist L., 1996, \apjs , 105, 19
  
\bibitem[]{Kay02} 
Kay S.T., Pearce F.R., Frenk C.S., Jenkins A., 2002, \mnras, 330, 113

\bibitem[]{Kle06}
Kleinheinrich M., et al., 2006, A\&A , 455, 441
  
\bibitem[]{Kra04} 
Kravtsov A.V., Berlind A.A., Wechsler R.H., Klypin A.A., Gottl\"ober
S., Allgood B., Primack J.R., 2004, \apj, 609, 35
  
\bibitem[]{Kly01} 
Klypin A., Kravtsov A.V., Bullock J.S., Primack J.R., 2001, \apj , 554, 903

\bibitem[]{Lu06}
Lu Y., Mo H.J., Katz N., Weinberg M.D., 2006, \mnras, 368, 1931

\bibitem[]{Lin06} 
Lin W.P., Jing Y.P., Mao S., Gao L., McCarthy I.G., preprint
  (astro-ph/0607555)

\bibitem[]{Man05a} 
Mandelbaum R., \etal , 2005a, \mnras , 361, 1287
  
\bibitem[]{Man05b} 
Mandelbaum R., Tasitsiomi A., Seljak U., Kravtsov A., Wechsler R.H., 
2005b, \mnras , 362, 1451
  
\bibitem[]{Man06a} 
Mandelbaum R., Seljak U., Kauffmann G.,  Hirata C.M.,
Brinkmann J., 2006a, \mnras, 368, 715

\bibitem[]{Man06b} 
Mandelbaum R., Seljak U., Cool R.J., Blanton M.,  Hirata C.M.,
Brinkmann J., 2006b, preprint (astro-ph/0605476)
  
\bibitem[]{McK01} 
McKay T.A., \etal , 2001, preprint (astro-ph/0108013)
  
\bibitem[]{McK02} 
McKay T.A., \etal , 2002, \apj , 571, L85

\bibitem[]{ME05.1}
Meneghetti M., Bartelmann M., Jenkins A., Frenk C., 2005,
preprint (astro-ph/0509323)

\bibitem[]{Mir91} 
Miralda-Escud\'e J., 1991, \apj, 370, 1
  
\bibitem[]{Moo99} 
Moore B., Quinn T., Governato F., Stadel J., Lake G., 1999,
\mnras , 310, 1147

\bibitem[]{nk97} 
Natarajan P., Kneib J.P., 1997, MNRAS, 287, 833

\bibitem[]{ns05} 
Natarajan P., Springel V., 2004, \apj , 617L, 13 

\bibitem[]{nds05}
Natarajan P., De Lucia G.,  Springel V., 2006 preprint (astro-ph/0604414)

\bibitem[]{nfw} 
Navarro J.F., Frenk C.S., White S.D.M., 1997, ApJ, 490, 493
  
\bibitem[]{nav04} 
Navarro J.F., \etal , 2004, \mnras , 349, 1039

\bibitem[]{Park05} 
Parker L.C., Hudson M.J., Carlberg R.G., Hoekstra H., 2005, \apj , 634, 806
  
\bibitem[]{Pow03} 
Power C., Navarro J.F., Jenkins A., Frenk C.S., White S.D.M., 
Springel V., Stadel J., Quinn T., 2003, \mnras , 338, 14

\bibitem[]{Reed05} 
Reed D., Governato F., Quinn T., Gardner J., Stadel J., Lake G., 
2005, \mnras , 359, 1537

\bibitem[]{SA04.1}
Sand D.J., Treu T., Smith G.P., Ellis R.S., 2004, \apj, 604, 88
  
\bibitem[]{sch05}
Schneider P., 2005, in: Kochanek C.S., Schneider P., Wambsganss J.: 
Gravitational Lensing: Strong, Weak \& Micro. G Meylan, P. Jetzer \& 
P. North (eds), Springer-Verlag: Berlin, P.273; preprint (astro-ph/0509252)
  
\bibitem[]{Sel00} 
Seljak U., 2000, \mnras , 318, 203
    
\bibitem[]{Sheldon04} 
Sheldon E.S., \etal , 2004, AJ, 127, 2544

\bibitem[]{Sim05}
Simon J.D., Bolatto A.D., Leroy A., Blitz L., Gates E.L., 2005, 
\apj, 621, 757

\bibitem{Smith01} 
Smith D.R., Bernstein G.M., Fischer P., Jarvis M. 2001, \apj, 551, 643
  
\bibitem[]{Som99} 
Somerville R.S., Primack J.R., 1999, \mnras , 310, 1087
  
\bibitem[]{Sper06} 
Spergel D.N., \etal , 2006, preprint (astro-ph/0603449)
  
\bibitem[]{Spr05a} 
Springel V. \etal , 2005, Nature, 435, 629
  
\bibitem[]{Spr05b} 
Springel V. 2005, \mnras , 364, 1105
  
\bibitem[]{Swa03}
Swaters R., Madore B.F., van den Bosch F.C., Balcells M., 2003, \apj,
583, 732

\bibitem[]{Tas04} 
Tasitsiomi A., Kravtsov A.V., Wechsler R.H., Primack J.R., 2004, 
\apj , 614, 533
  
\bibitem[]{Tink05} 
Tinker J.L., Weinberg D.H., Zheng Z., Zehavi I., 2005, \apj, 631, 41
  
\bibitem{Tyson84} 
Tyson J.A., Valdes F., Jarvis J.F., Mills A.P. Jr.  1984, \apj , 281L, 59
  
\bibitem[]{vale04} 
Vale A., Ostriker J.P., 2004, \mnras , 353, 189
  
\bibitem[]{vale06} 
Vale A., Ostriker J.P., 2006, \mnras , in press, preprint (astro-ph/0511816)
  
\bibitem[]{Bosch02} 
van den Bosch F.C., 2002, \mnras , 332, 456

\bibitem[]{BBDB} 
van den Bosch F.C., Robertson B.E., Dalcanton J.J., de Blok W.J.G.,
2000, \aj, 119, 1579
  
\bibitem[]{BYM03} 
van den Bosch F.C., Yang X., Mo H.J., 2003, \mnras , 340, 771
    
\bibitem[]{BYM04} 
van den Bosch F.C., Norberg P., Mo H.J., Yang X., 2004, \mnras, 352, 1302
  
\bibitem[]{vdB05} 
van den Bosch F.C., Tormen G., Giocoli C., 2005a, \mnras , 359, 1029
  
\bibitem[]{BWYMLJ} 
van den Bosch F.C., Weinmann S.M., Yang X., Mo H.J., Li C., Jing Y.P., 
2005b, \mnras , 351, 1203

\bibitem[]{BYMN} 
van den Bosch F.C., Yang X., Mo H.J., Norberg P., 2005c, \mnras , 356, 1233

\bibitem[]{Wech02} 
Wechsler R.H., Bullock J.S., Primack J.R., Kravtsov A.V., Dekel A., 
2002, \apj , 568, 52

\bibitem[]{Wein04} 
Weinberg D.H., Dav'e R., Katz N., Hernquist L., 2004, \apj, 601, 1
  
\bibitem[]{Wei06} 
Weinmann S.M., van den Bosch F.C., Yang X., Mo H.J., 2006, \mnras, 366, 2 

\bibitem[]{Whit01} 
White M., 2001, \mnras , 321, 1
  
\bibitem[]{White91} 
White S.D.M., Frenk C., 1991, \apj , 379, 52
  
\bibitem{Wilson01} 
Wilson G., Kaiser N., Luppino G.A., Cowie L.L., 2001, ApJ, 555, 572

\bibitem{Wright00}
Wright C.O., Brainerd T.G., 2000, ApJ, 534, 34
  
\bibitem[]{YMB03a} 
Yang X., Mo H.J., Kauffmann G., Chu Y.Q., 2003a, \mnras , 339, 387
  
\bibitem[]{YMB03} 
Yang X., Mo H.J., van den Bosch F.C., 2003b, \mnras , 339, 1057
  
\bibitem[]{YMJBC} 
Yang X., Mo H.J., Jing Y.P., van den Bosch F.C., Chu Y., 2004, 
\mnras, 350, 1153
  
\bibitem[]{Y05a} 
Yang X., Mo H.J., van den Bosch F.C., Jing Y.P., 2005a, \mnras, 356, 1293
  
\bibitem[]{Y05b} 
Yang X., Mo H.J., Jing Y.P., van den Bosch F.C., 2005b, \mnras , 358, 217

\bibitem[]{Yoo05}
Yoo J., Tinker J.L., Weinberg D.H., Zheng Z., Katz N., Dav\'e R.,
2005, \apj , in press, preprint (astro-ph/0511580)

\bibitem[]{zhao03a} 
Zhao D.H., Mo H.J., Jing Y.P., B\"orner G., 2003a, \mnras, 339, 12
  
\bibitem[]{zhao03b} 
Zhao D.H., Jing Y.P., Mo H.J., B\"orner G., 2003b, \apj, 597L, 227
  
\bibitem[]{zhao96b} 
Zhao H., 1996, \mnras , 278, 488

 \bibitem[]{Zhe05} 
Zheng Z., et al., 2005, \apj , 633, 791

\end{thebibliography}
\end{document}